%% file: main.tex
\documentclass{article}

\usepackage{microtype}
\usepackage{graphicx}
\usepackage{subfigure}
\usepackage{booktabs} %

\usepackage{amsmath,amsfonts}
\usepackage{physics}
\usepackage{stfloats}%
\usepackage{cuted}
\usepackage[nolist]{acronym}

\usepackage{hyperref}
\usepackage{pifont}

\usepackage{comment}

\DeclareMathOperator{\MMD}{MMD}
\DeclareMathOperator{\MMDH}{\MMD_{\mathbb{H}}}
\newcommand{\joint}{\mathbb{P}_Z}
\newcommand{\factor}{\mathbb{P}_{\bar{Z}}}

\usepackage{varwidth}
\usepackage{tikz}
\usepackage{pgf}
\usepackage{pgfplots}
\usepackage{pgfplotstable}
\usetikzlibrary{positioning,backgrounds,fit,calc,patterns}
\pgfplotsset{compat=1.3}
\usetikzlibrary{shapes.geometric}
\usetikzlibrary{shapes.arrows}
\usepackage{array}
\usepackage[many]{tcolorbox}

\usepackage[accepted]{icml2024}

\icmltitlerunning{An Independence-promoting Loss for Music Generation with Language Models}

\begin{document}

\twocolumn[
\icmltitle{An Independence-promoting Loss for \\ Music Generation with Language Models}

\icmlsetsymbol{equal}{*}
\icmlsetsymbol{metatmp}{$\dagger$}

\begin{icmlauthorlist}
\icmlauthor{Jean-Marie Lemercier}{uhh,metatmp,equal}
\icmlauthor{Simon Rouard}{ircam,meta,equal}
\icmlauthor{Jade Copet}{meta}
\icmlauthor{Yossi Adi}{meta,jerus}
\icmlauthor{Alexandre D\'efossez}{kyutai,metatmp}
\end{icmlauthorlist}

\icmlaffiliation{uhh}{Universit\"at Hamburg}
\icmlaffiliation{ircam}{IRCAM}
\icmlaffiliation{meta}{Meta AI}
\icmlaffiliation{kyutai}{Kyutai $^\dagger$Work done while at Meta AI}
\icmlaffiliation{jerus}{The Hebrew University of Jerusalem}

\icmlcorrespondingauthor{Jean-Marie Lemercier}{jeanmarie.lemercier@uni-hamburg.de}

\icmlkeywords{Language Models, Audio Generation, Music Generation, Information Theory, Independence}

\vskip 0.3in
]

\printAffiliationsAndNotice{\icmlEqualContribution} %

\begin{acronym}
\acro{stft}[STFT]{short-time Fourier transform}
\acro{istft}[iSTFT]{inverse short-time Fourier transform}
\acro{dnn}[DNN]{deep neural network}
\acro{pesq}[PESQ]{Perceptual Evaluation of Speech Quality}
\acro{polqa}[POLQA]{perceptual objectve listening quality analysis}
\acro{wpe}[WPE]{weighted prediction error}
\acro{psd}[PSD]{power spectral density}
\acro{rir}[RIR]{room impulse response}
\acro{snr}[SNR]{signal-to-noise ratio}
\acro{lstm}[LSTM]{long short-term memory}
\acro{polqa}[POLQA]{Perceptual Objectve Listening Quality Analysis}
\acro{sdr}[SDR]{signal-to-distortion ratio}
\acro{estoi}[ESTOI]{Extended Short-Term Objective Intelligibility}
\acro{elr}[ELR]{early-to-late reverberation ratio}
\acro{tcn}[TCN]{temporal convolutional network}
\acro{rls}[RLS]{recursive least squares}
\acro{asr}[ASR]{automatic speech recognition}
\acro{ha}[HA]{hearing aid}
\acro{ci}[CI]{cochlear implant}
\acro{mac}[MAC]{multiply-and-accumulate}
\acro{vae}[VAE]{variational auto-encoder}
\acro{gan}[GAN]{generative adversarial network}
\acro{tf}[T-F]{time-frequency}
\acro{sde}[SDE]{stochastic differential equation}
\acro{ode}[ODE]{ordinary differential equation}
\acro{drr}[DRR]{direct to reverberant ratio}
\acro{lsd}[LSD]{log spectral distance}
\acro{sisdr}[SI-SDR]{scale-invariant signal to distortion ratio}
\acro{mos}[MOS]{mean opinion score}
\acro{map}[MAP]{maximum a posteriori}
\acro{rtf}[RTF]{real-time factor}
\acro{ema}[EMA]{exponential moving average}
\acro{rkhs}[RKHS]{reproducible kernel Hilbert space}
\acro{mmd}[MMD]{maximum mean discrepancy}
\acro{ica}[ICA]{independence component analysis}
\end{acronym}

\begin{abstract}
Music generation schemes using language modeling rely on a vocabulary of audio tokens, generally provided as codes in a discrete latent space learnt by an auto-encoder.
Multi-stage quantizers are often employed to produce these tokens, therefore the decoding strategy used for token prediction must be adapted to account for multiple codebooks: 
either it should model the joint distribution over all codebooks, or fit the product of the codebook marginal distributions.
Modelling the joint distribution requires a costly increase in the number of auto-regressive steps, while fitting the product of the marginals yields an inexact model unless the codebooks are mutually independent.
In this work, we introduce an independence-promoting loss to regularize the auto-encoder used as the tokenizer in language models for music generation. 
The proposed loss is a proxy for mutual information based on the maximum mean discrepancy principle, applied in reproducible kernel Hilbert spaces. 
Our criterion is simple to implement and train, and it is generalizable to other multi-stream codecs. 
We show that it reduces the statistical dependence between codebooks during auto-encoding.
This leads to an increase in the generated music quality when modelling the product of the marginal distributions, while generating audio much faster than the joint distribution model.
\end{abstract}

\section{Introduction}
\input{sections/intro}

\section{Background}
\input{sections/background}

\section{Method} \label{sec:method}
\input{sections/method.tex}

\section{Experiments}
\input{sections/xp.tex}

\section{Results}
\input{sections/results.tex}

\section{Conclusion}
We presented an independence-proxy loss for regularizing discrete latent representations used as tokens in music generation language models.
We showed that the proposed method outperforms our baseline and other state-of-the-art music generation models, without adding parameters nor increasing the inference time compared to the baseline.
We performed an analysis of the propose criterion, showing its correlation with total correlation of the codes and investigating the effects of adapting the criterion to the decoding strategy used in further language modelling.
We also demonstrated that the proposed criterion can be easily plugged into other multi-stream codecs, and more generally we would argue that is is a reasonable independence optimization criterion for other applications than music generation.

\section*{Impact Statement}
Large scale generative models boast high expression capabilities, which raises questions regarding ethics and societal consequences of their use.
In particular, text-to-music generative models can constitute an unfair competition for musicians (and artists and creators in general). This is a societal issue that has not been solved yet and demands serious regulatory investigation. We try and make our research as open and accessible as possible, ensuring that the involved parties, both amateurs and professional, have equal access to the developed methods. 
Another potential bias towards individuals resides in the large proportion of Western music (and in particular pop instrumental and electronic music) of the data used to train our model, which resents a lack of diversity.
However, the somewhat reasonable size of the model presented in this paper and the low number of auto-regressive steps used for inference should encourage reproducibility of our method for new data sources.

\bibliography{main}
\bibliographystyle{icml2024}

\newpage
\appendix
\input{sections/appendix}

\end{document}

%% file: sections/intro.tex
Generative models are being increasingly used to produce multimedia content such as e.g. image \cite{rombach2022highresolution},
text \cite{brown2020language}, speech \cite{oord2016wavenet, kong2020hifigan, kong2021diffwave} or
audio \cite{borsos2023audiolm, agostinelli2023musiclm, yang2023diffsound, kreuk2023audiogen}.
These models rely on artificial neural networks parameterizing approaches such as
generative adversarial networks
\cite{goodfellow2014gan}, diffusion models \cite{ho2020denoising, song2019generative} or transformer-based language models \cite{radford2019language, vaswani2023attention}.
We focus here on the task of generating music based on a text prompt. Music signals occupy the full frequency spectrum (unlike speech) and can be very long sequences (unlike most images), making the generation task arduous.
Text-to-music language models \cite{agostinelli2023musiclm, kreuk2023audiogen, copet2023musicgen, borsos2023audiolm}
try to model the distribution of a vocabulary of discrete units i.e. tokens. The audio tokens are often generated by a multi-stage quantizer operating in the latent space learnt by a neural compression model \cite{defossez2023high, zeghidour2021soundstream}.
As the quantizer uses a distinct codebook for each stage, the language model decoding strategy must be adapted to model either the joint distribution over all codebooks, or the factorization of codebook marginal distributions.
On the one hand, modelling the joint distribution requires either using an impractically large vocabulary size, or multiplying the number of auto-regressive timesteps by the number of codebooks.
On the other hand, modelling the factorized distribution significantly facilitates the training of the language model and speeds inference up, but only provides an approximation of the true model.
Several strategies for modelling the factorized distribution have been proposed \cite{wang2023neural, kharitonov2022textfree, kreuk2023audiogen, copet2023musicgen}
yielding satisfying results. However, we argue that these solutions do not directly address the issue, which is that the factorized distribution is equivalent to the full joint distribution \textit{only if the codebooks are mutually independent}.

In this work, we propose to introduce an independence constraint between codebooks, in the form of an auxiliary objective for training the auto-encoder used as the tokenizer for the language model.
Instead of leveraging adversarial training as in \cite{belghazi_mine_2021, brakel2017learning}, we propose to use a proxy for mutual information based on the maximum mean discrepancy \cite{mmd},
which solves a dual formulation of earth mover optimization in Gaussian reproducible kernel Hilbert spaces.
We conduct experiments on music generation, and run ablations with respect to our independence-promoting loss configurations.

We make the following contributions:
\begin{itemize}
    \item We show that the maximum mean discrepancy in reproducible kernel Hilbert spaces is a reasonable proxy for independence, since optimizing our criterion leads to a reduction of mutual information between codebooks during auto-encoding.

    \item We propose a modified version of our loss that matches the decoding strategy used for token prediction. When applied to the ``delay'' strategy proposed in \cite{kharitonov2022textfree}, we obtain the best performance across all our models.
    
    \item We show that objective and subjective music generation quality scores favour the language model whose tokenizer was trained with the proposed independence loss in comparison to other baselines. 
    Our resulting model has the same amount of parameters and generation speed as the baseline not using our proposed criterion.
    Our approach enables to generate audio at the same frame rate as the auto-encoder, which is much faster than the joint distribution model and has similar generation quality.

\end{itemize}
Please visit our companion website\footnote{\href{https://jmlemercier.github.io/encodec-mmd.github.io}{encodec-mmd.github.io}} for audio examples, support with code, etc.

%% file: sections/background.tex
\subsection{Quantization}
\label{sec:quantization}

Quantization is a discretization method aiming at reducing the bitrate used to encode information, which is a major challenge in low-resource communications. Quantization is also  used in machine learning, typically to reduce the memory and computational footprints of \acp{dnn} on embedded devices. More recently, quantizers were used to produce a vocabulary of discrete units for language models learning the distribution of originally continuous signals such as e.g. images or audio.
Quantization schemes can be categorized in two classes: scalar and vector quantization. Scalar quantization discretizes each dimension of the considered signal, rounding the current value to the closest bin on a quantization grid.
Vector quantization (VQ) \cite{Gray1984VectorQ} encodes signals as entries (or \textit{codes}) in a multi-dimensional codebook.
Concretely, VQ learns a codebook $\mathcal{C}$ with $M$ vectors of dimension $N$ and at inference, it performs a nearest neighbour search in the codebook space to find the right code for the input signal.

Multi-stage vector quantizers \cite{juang1982multistage, vasuki2006review} use multiple codebooks with reasonable size, which increases codebook utilization compared to having one large codebook. This is one of the keys to the success of these structured quantizers, which achieve a good trade-off between computational complexity and coding efficiency.
Residual vector quantization (RVQ) \cite{zeghidour2021soundstream} is a multi-stage vector quantization scheme that introduces $K$ codebooks. At each stage $k \in \{1, \dots, K \}$, the residual of the previous stage is quantized with the codebook $\mathcal{C}^{(k)}$ and the residual for the next stage is obtained by subtracting the resulting code from the previous residual.
The codes exhibit a natural hierarchical, coarse-to-fine structure, as most of the information is contained in the first few codebooks.

\subsection{Independence of Random Variables}
\label{sec:mutual_information}

Reliably measuring statistical dependence between random variables is a wide-spread topic in the machine learning literature \cite{betavae, burgess2018understanding, brakel2017learning, hyvarinen_nonlinear_2023, belghazi_mine_2021}.
Let \{$Z_1, \dots, Z_K\} $ be a family of vector random variables in $\mathbb{R}^N$.
It is an independent family if and only if the joint distribution, denoted as $\mathbb{P}_Z$, and the product of the marginal distributions denoted as $\mathbb{P}_{\bar{Z}}$ (or \textit{factorized} distribution) coincide.
This is equivalent to saying that the joint probability density function
can be factorized into the product of the marginal probability density functions, i.e.
$\forall J \leq K , \, \forall (k_1, \dots k_{J}) \in \{ 1, \dots, K \}^{J}$ with ${i} \neq {j} \Rightarrow k_i \neq k_j$ and $\forall (z_{k_1}, \dots z_{k_{J}}) \in \mathbb{R}^{N \times J}$:
\begin{equation}
    p_{ Z_{k_1}, \dots, Z_{k_J} }( z_{k_1}, \dots, z_{k_J} )
    = \Pi_{j=1}^J p_{Z_{k_j}}( z_{k_j} ). 
\end{equation}
where $p_X$ is the probability density function of the random variable $X$.
Independence between variables can be exactly measured via the mutual information $\mathcal{I}(Z_1, \dots Z_K)$, which equals the Kullback-Leibler divergence between the joint distribution $\mathbb{P}_Z$ and the factorized distribution $\mathbb{P}_{\bar{Z}}$. This instance of mutual information is called \textit{total correlation}, and can also be expressed in terms of entropies:
\begin{align}
\mathcal{I}(Z_1, \dots Z_K) 
&= \mathrm{D}_\mathrm{KL} \left( \mathbb{P}_Z || \mathbb{P}_{\bar{Z}} \right) \\  \label{eq:mi}
&= \mathcal{H}(Z_1, \dots, Z_K) - \sum_{k=1}^K \mathcal{H}(Z_k),
\end{align}
where $\mathcal{H}(X)$ measures the entropy of the random variable $X$.
While a closed-form computation of the total correlation is available through \eqref{eq:mi}, this requires either exact knowledge of the distributions,
or approximate knowledge through histogram estimation.
We will eliminate the first option since we do not posit distributional assumptions as in e.g. the \ac{vae} case \cite{kingma2014vae, betavae}.
Estimating the histogram of the marginal variables $Z_i$ might be possible most of the time. However, estimating the histogram of the joint variable $(Z_1, \dots, Z_K)$ is a tedious operation as it requires an immense sample size.
Another poor property of histograms is that their computation is not differentiable.

For the reasons listed above, we should resort to proxies to force the independence of random variables.
Several independence proxies have already been proposed in the literature \cite{belghazi_mine_2021, brakel2017learning, ddica}. However, these often rely on adversarial training, which is known to significantly increase the training difficulty \cite{goodfellow2014gan}. For instance \cite{belghazi_mine_2021} optimize a dual formulation of the Kullback-Leibler divergence through adversarial training of neural estimators. A similar paradigm was already explored for non-linear \ac{ica} \cite{hyvarinen_nonlinear_2023}, where a neural network was trained to discriminate between samples from the joint distribution and samples from the factorized distribution \cite{brakel2017learning}. A Jensen-Shannon divergence objective is then formulated and optimized using the estimated joint-to-factorized probability ratio
\cite{huszar_blogpost}.

Aside the Kullback-Leibler and Jensen-Shannon divergences, another convenient distance between probability distributions is the earth mover distance, defined as:
\begin{equation} \label{eq:mmd}
    W_2(\mathbb{P}_Z || \mathbb{P}_{\bar{Z}}) = \inf_{\pi \in \Pi(\mathbb{P}_Z, \mathbb{P}_{\bar{Z}})} \mathbb{E}_{(Z, \bar{Z}) \sim \pi} \norm{Z - \bar{Z}}_2,
\end{equation}
where $\Pi(\mathbb{P}_Z, \mathbb{P}_{\bar{Z}})$ denotes the ensemble of all distributions whose marginals are $\mathbb{P}_Z$ and $ \mathbb{P}_{\bar{Z}}$.
Given the Kantorovic-Rubinstein duality \cite{villani2009ot}, the earth mover distance coincides with the \ac{mmd} \cite{mmd} defined as a simpler optimization problem over real-valued $1$-Lipschitz functions:
\begin{align} \label{eq:mmd}
    \MMD(\mathbb{P}_Z || \mathbb{P}_{\bar{Z}}) &= W_2(\mathbb{P}_Z || \mathbb{P}_{\bar{Z}}) \nonumber \\ 
    &= \sup_{h, \norm{h} \leq 1} \mathbb{E}_{Z \sim \mathbb{P}_Z}[h(Z))] - \mathbb{E}_{\bar{Z} \sim \mathbb{P}_{\bar{Z}}}[h(\bar{Z})].
\end{align}
Since MMD is equivalent to the earth mover distance, if $\MMD(\mathbb{P}_Z || \mathbb{P}_{\bar{Z}}) = 0$ then the joint distribution $\mathbb{P}_Z$ and the factorized distribution $\mathbb{P}_{\bar{Z}}$ are equal and therefore the family $\{ Z_1, \dots, Z_K \}$ is independent.

One could use a neural network to parameterize the function $h$ and train it with an adversarial loss, which would resemble the aforementioned works \cite{belghazi_mine_2021, brakel2017learning}. This was applied in \cite{arjosky2014wgan}, although for density estimation in \acp{gan} rather than independence optimization.
However, \cite{mmd} highlight a remarkable property of the MMD by taking the set of functions $h$ to be the unit ball in an \ac{rkhs} $\mathbb{H}$.

Let $X \in \mathbb{R}^{N \times J}$: an \textit{evaluation operator} $\delta_X : \mathbb{H} \rightarrow \mathbb{R}$ associates $h \in \mathbb{R}$ to its evaluation $h(X) \in \mathbb{R}$.
The Riesz representation theorem guarantees that for each continuous evaluation operator $\delta_X$, there exists a \textit{feature mapping} $\phi(X) \in \mathbb{H}$, such that $\forall h \in \mathbb{H}, \delta_X(h) := h(X) = \langle h, \phi(X) \rangle_{\mathbb{H}}$. 
A core property of \acp{rkhs} is that they are equipped with a kernel function $k: \mathbb{R}^{N \times J} \times \mathbb{R}^{N \times J} \rightarrow \mathbb{R}$, such that dot products between features can be conveniently computed as $\langle \phi(X), \phi(Y) \rangle_\mathbb{H} = k(X, Y)$.
It can be then shown that a lower-bound of the MMD in \eqref{eq:mmd} can be obtained as a combination of kernel computations: 
\begin{align} \label{eq:mmd_kernel}
    \MMDH(\mathbb{P}_Z || \mathbb{P}_{\bar{Z}}) & = \, \, \mathbb{E}_{Z_1 \sim \mathbb{P}_Z} \mathbb{E}_{Z_2 \sim \mathbb{P}_Z} k(Z_1, Z_2) \nonumber \\
    &+ \, \, \,
    \mathbb{E}_{\bar{Z}_1 \sim \mathbb{P}_{\bar{Z}}} \mathbb{E}_{\bar{Z}_2 \sim \mathbb{P}_{\bar{Z}}} k(\bar{Z}_1, \bar{Z}_2) \\
    &- 2 \mathbb{E}_{Z_1 \sim \mathbb{P}_Z} \mathbb{E}_{\bar{Z}_2 \sim \mathbb{P}_{\bar{Z}}} k(Z_1, \bar{Z}_2) \nonumber \\
    &\leq \MMD(\mathbb{P}_Z || \mathbb{P}_{\bar{Z}}). \nonumber
\end{align}
The proof is let to appendix \ref{appendix:proof}.
An important property of $\MMDH$ is that if $\mathbb{H}$ is a \textit{universal} RKHS, then $\MMDH(\joint||\factor)=0 \iff \joint=\factor$ \cite{mmd}. 
This shows that if we achieve optimality for our lower-bound $\MMDH$ using a universal RKHS, we actually obtain an independent representation. 
A RKHS $\mathbb{H}$ is said universal if it is dense in the space of functions $h: \mathbb{R}^{N \times J} \mapsto \mathbb{R}$. In particular, RKHSs with Gaussian kernels are universal. 

Our proposed proxy can easily be computed with batch estimators and does not require adversarial training.
Another kernel-based estimator was presented in \cite{ddica, yu_measuring_2021}. However, it requires a singular-value decomposition of the kernel matrices $k(Z_1, Z_2)$ which is sensitive to numerical errors, produces gradients with high variance and is costly for high-dimensional data.

\subsection{Audio Generation with Language Models}
\label{sec:lm}

Language modelling using auto-regressive Transformer-style architectures \cite{vaswani2023attention} has been central in audio generation lately \cite{dhariwal2020jukebox, borsos2023audiolm, wang2023neural, agostinelli2023musiclm, kreuk2023audiogen, copet2023musicgen}.
These approaches typically consist of two modules.
The first is a neural audio compression model such as e.g. \cite{zeghidour2021soundstream, defossez2023high} that takes as input the raw audio $X \in \mathbb{R}^{L}$ with $L$ the sequence length.
The encoder part of this codec transforms $X$ into a discrete token sequence with codebook indexes $Q \in \{ 1, \dots, M \}^{T \times K}$ and corresponding codes $Z \in \mathbb{R}^{T \times K \times N}$, where $T$ is the reduced time length obtained via the encoder strides, $K$ is the number of codebooks, $M$ is the codebook size and $N$ is the codebook dimension.
The second module is an autoregressive Transformer-decoder language model operating in the space of discrete audio tokens. Given a textual conditioning $C$ provided by a pre-trained text encoder, the language model $f_\theta$ predicts the distribution of a sequence of tokens $Z$ auto-regressively as $f_\theta(Z^{(t)} | \, C, Z^{(1)}, \dots, Z^{(t-1)})$.
Finally, the acoustic tokens generated by the language model are provided to the audio decoder to synthesize the final waveform.

Because VQ-based audio codecs typically use multiple codebooks for optimal compression, the usual single-stream decoding strategy of language models needs to be adapted.
The token sequence can be for instance flattened, and the transformer then predicts the codebooks sequentially. Theoretically, this leads to modelling the joint distribution of codebooks $\mathbb{P}_Z$ \cite{copet2023musicgen}. However, this approach yields high computational complexity as the frame rate is multiplied by the number of codebooks $K$ compared to the auto-encoder.

Another solution is to decode the distributions of each codebook independently and thus modelling the factorized distribution $\mathbb{P}_{\bar{Z}}$ conditionally to the past tokens $\{ Z^{(1)}, \dots, Z^{(t-1)} \}$. However, this approach is only equivalent to the exact model of the joint distribution $\mathbb{P}_Z$ if the codes of each codebook are mutually independent, conditionally to the past codes.
Using the concepts introduced in \ref{sec:mutual_information}, this means the family $\{ Z_1^{(t)}, \dots Z_K^{(t)} \}$ should be independent, conditionally to $\{ Z^{(1)}, \dots, Z^{(t-1)} \}$.
As $t$ increases, errors due to statistical dependence between codes may compound and cause the model to diverge from the true distribution.
However, this method preserves the original codec frame rate, significantly accelerating training and inference.

Several alternative decoding strategies have been introduced: \cite{wang2023neural} propose to fully model the distribution of the first codebook, then to learn the factorized distribution over the remaining codebooks, while \cite{borsos2023audiolm, agostinelli2023musiclm} model the first four codebooks with a first decoder, then the remaining eight codebooks with a second decoder.
\cite{kharitonov2022textfree} introduce a delay between codebooks for multi-stream language modeling, as an alternative to simply modelling all codebooks in parallel. This was used for audio and music generation in \cite{kreuk2023audiogen} and \cite{copet2023musicgen}, respectively.

We propose instead to address the issue of statistical dependence between codes, so that we can reduce the modelling error but keep the inference time low when modelling the factorized distribution. This is the objective of the next section, where we present our independence promoting loss.

%% file: sections/method.tex
\input{figures/diagram}
\definecolor{figc0}{HTML}{ff5733}

We introduce here our proposed objective loss for promoting independence between codebooks.
Using the maximum mean discrepancy framework presented in Section~\ref{sec:mutual_information}, we choose a reproducible kernel Hilbert space $\mathbb{H}$ equipped with a kernel $k( \cdot,  \cdot )$.
We do not operate in a variational framework, and consequently do not posit assumptions as to how the codes are distributed in the latent space. Therefore, we need to work with empirical estimators.
An unbiased empirical estimator for the MMD lower-bound between samples $\{ Z_i \}_{i=1}^B$ and $\{ \bar{Z}_i \}_{i=1}^B$ is obtained from \eqref{eq:mmd_kernel}:
\begin{align} \label{eq:mmd_estimator}
    \MMD_{\mathbb{H}}(\mathbb{P}_Z || \mathbb{P}_{\bar{Z}})
    &= \frac{1}{B(B-1)} \displaystyle{\sum_{i=1}^B \sum_{j\neq i}} k(Z_i, Z_j) \nonumber \\
    &+ \frac{1}{B(B-1)} \displaystyle{\sum_{j=1}^B \sum_{j\neq i}} k(\bar{Z}_i, \bar{Z}_j) \nonumber \\
    &- \frac{2}{B^2} \displaystyle{\sum_{i=1}^B \sum_{j=1}^B} k(Z_i, \bar{Z}_j),
\end{align}
where $B$ is the sample size and $i, j$ are indexes of samples in the batch.

Given a batch of samples $\{ Z_i \}_{i=1}^B$
of the joint distribution $\mathbb{P}_Z$ obtained via encoding and quantization, we use the same batch shuffling strategy as \cite{brakel2017learning} to obtain samples $\{ \bar{Z}_i \}_{i=1}^B$ 
of the factorized distribution $\mathbb{P}_{\bar{Z}}$.
For each codebook, we randomly shuffle the corresponding codes along the batch dimension, which was shown to effectively approximate samples of the factorized distribution $\mathbb{P}_{\bar{Z}}$ for sufficiently large sample sizes $B$. As explained further in the experiments section, we choose the sample size to be as large as possible to reduce both the variance of the empirical $\MMDH$ estimator and the reshuffling algorithm.
The independence loss \textcolor{figc0}{$\mathcal{L}_\mathrm{inde}$} is then obtained by computing the empirical $\MMDH$ estimator between samples from the joint and approximate factorized distributions, as summarized in Algorithm~\ref{alg:training}.
Note that by promoting independence between codeboks through optimization of $\MMDH$, we actually achieve more than the weaker \textit{conditional} independence required by our decoding strategies to obtain exact modeling.
Designing a conditional independence objective is not explored here.

\input{figures/algo.tex}

This version of the proposed auxiliary loss promotes independence between the codes corresponding to encoded frames with similar frame index. This is optimal when adopting a parallel decoding strategy, effectively
modelling the factorized distribution $\mathbb{P}_{\bar{Z}}$.
We propose to extend our independence-promoting by applying the ``delay'' strategy proposed in \cite{kharitonov2022textfree} 
to the codes before computing the $\MMDH$ estimator, effectively promoting independence between time-delayed codes $\{ Z_k^{(.-k+1)} \}_{k=1}^K$, as this will be our token decoding strategy for language modelling.
The same could be done for other decoding strategies such as e.g. Vall-E \cite{wang2023neural}. 
A diagram of the whole framework is displayed in Figure~\ref{fig:diagram}.

%% file: figures/diagram.tex
\tikzstyle{mycircle} = [circle, draw, fill=white, inner sep=0pt, minimum size=40pt]
\tikzstyle{mysquare} = [rectangle, draw, fill=white, inner sep=0pt, minimum size=30pt]
\tikzstyle{myrectangle3} = [rectangle, draw, fill=white, inner sep=0pt, minimum width=2cm, minimum height=1cm, align=center]
\tikzstyle{mybranch} = [circle, draw, fill=black, inner sep=0pt, minimum size=3pt]
\tikzstyle{mylittlesquare} = [rectangle, draw, fill=white, inner sep=0pt, minimum size=15pt]
\tikzstyle{mytrapeze} = [trapezium, trapezium angle=70, minimum height=1.35cm, draw]

\tikzstyle{sum} = [
  circle,
  draw,
  minimum size=12pt,
  append after command={
    \pgfextra{
      \draw (\tikzlastnode.north) -- (\tikzlastnode.south);
      \draw (\tikzlastnode.west) -- (\tikzlastnode.east);
    }
  },
]

\newcommand{\scale}{1.0}
\newcommand{\sepblocks}{0.5cm}
\newcommand{\sepblockslm}{0.2cm}
\newcommand{\sepcodes}{0.1cm}

\definecolor{figblue}{HTML}{154c79}
\definecolor{figdarkergreen}{HTML}{2ba481}
\definecolor{figc0}{HTML}{ff5733}
\definecolor{figc1}{HTML}{ffaf33}
\definecolor{figc2}{HTML}{ffe933}
\definecolor{figc3}{HTML}{a5ff33}
\definecolor{figc4}{HTML}{33ff86}
\definecolor{figc5}{HTML}{33f9ff}
\definecolor{figc6}{HTML}{33a5ff}
\definecolor{figc7}{HTML}{7d33ff}

\begin{figure*}[h]

\centering
\scalebox{\scale}{
\begin{tikzpicture}

\coordinate (pos_wav_input) at (1.5, 10);

\node[mycircle, dashed, fill overzoom image={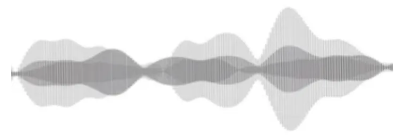}] (input) at (pos_wav_input) {};
\node[mytrapeze, align=center, right=\sepblocks of input, shape border rotate=270] (encoder) {Audio\\Encoder};
\node[mysquare, right=\sepblocks of encoder] (quantizer) {RVQ};
\node[mytrapeze, right=\sepblocks of quantizer, align=center, shape border rotate=90] (decoder) {Audio\\Decoder};
\node[mycircle, dashed, fill overzoom image={figures/waveform_meta.png}, right=\sepblocks of decoder] (output) {};

\draw[-, dashed] (input) to (encoder);
\draw[-] (encoder) to (quantizer);
\draw[-] (quantizer) to (decoder);
\draw[-, dashed] (decoder) to (output);

\node[above right=1.2cm and 0.5cm of quantizer.center] (l_inde) {$\textcolor{figc0}{\mathcal{L}_\mathrm{inde}}$};
\node[above left=1.2cm and 0.5cm of quantizer.center] (l_codebook) {$\textcolor{figblue}{\mathcal{L}_\mathrm{codebook}}$};
\node[above left=1.2cm and 0.5cm of encoder.center] (l_commit) {$\textcolor{figblue}{\mathcal{L}_\mathrm{commit}}$};
\node[above right=1.2cm and 0.5cm of output.center] (l_reconstruction) {$\textcolor{figblue}{\mathcal{L}_\mathrm{rec}}$};
\node[below right=1.2cm and 0.5cm of output.center] (l_adversarial) {$\textcolor{figblue}{\mathcal{L}_\mathrm{adv}}$};

\draw[-, thick, loosely dotted, figblue] (quantizer.north west) to (l_codebook);
\draw[-, thick, loosely dotted, figc0] (quantizer.north east) to (l_inde);
\draw[-, thick, loosely dotted, figblue] (encoder) to (l_commit);
\draw[-, thick, loosely dotted, figblue] (output) to (l_reconstruction);
\draw[-, thick, loosely dotted, figblue] (output) to (l_adversarial);

\coordinate (pos_codebooks) at (-3.2, 5.4);

\node[myrectangle3, dashed, minimum width=3.9cm, minimum height=2.67cm, anchor=south west] (codebooks) at (pos_codebooks) {};

\node[mylittlesquare, fill=figc0, above right=\sepcodes and \sepcodes of codebooks.south west] (t00) {$Z_1^1$};
\node[mylittlesquare, above=\sepcodes of t00] (t01) {};
\node[mylittlesquare, above=\sepcodes of t01] (t02) {};
\node[mylittlesquare, above=\sepcodes of t02] (t03) {};

\node[mylittlesquare, fill=figc1, right=\sepcodes of t00] (t10) {$Z_1^2$};
\node[mylittlesquare, fill=figc0, above=\sepcodes of t10] (t11) {$Z_2^1$};
\node[mylittlesquare, above=\sepcodes of t11] (t12) {};
\node[mylittlesquare, above=\sepcodes of t12] (t13) {};

\node[mylittlesquare, fill=figc2, right=\sepcodes of t10] (t20) {$Z_1^3$};
\node[mylittlesquare, fill=figc1, above=\sepcodes of t20] (t21) {$Z_2^2$};
\node[mylittlesquare, fill=figc0, above=\sepcodes of t21] (t22) {$Z_3^1$};
\node[mylittlesquare, above=\sepcodes of t22] (t23) {};

\node[mylittlesquare, fill=figc3, right=\sepcodes of t20] (t30) {$Z_1^4$};
\node[mylittlesquare, fill=figc2, above=\sepcodes of t30] (t31) {$Z_2^3$};
\node[mylittlesquare, fill=figc1, above=\sepcodes of t31] (t32) {$Z_3^2$};
\node[mylittlesquare, fill=figc0, above=\sepcodes of t32] (t33) {$Z_4^1$};

\node[mylittlesquare, fill=figc4, right=\sepcodes of t30] (t40) {$Z_1^5$};
\node[mylittlesquare, fill=figc3, above=\sepcodes of t40] (t41) {$Z_2^4$};
\node[mylittlesquare, fill=figc2, above=\sepcodes of t41] (t42) {$Z_3^3$};
\node[mylittlesquare, fill=figc1, above=\sepcodes of t42] (t43) {$Z_4^2$};

\node[mylittlesquare, fill=figc5, right=\sepcodes of t40] (t50) {$Z_1^6$};
\node[mylittlesquare, fill=figc4, above=\sepcodes of t50] (t51) {$Z_2^5$};
\node[mylittlesquare, fill=figc3, above=\sepcodes of t51] (t52) {$Z_3^4$};
\node[mylittlesquare, fill=figc2, above=\sepcodes of t52] (t53) {$Z_4^3$};

\node[myrectangle3, minimum width=20pt, minimum height=2.67cm, dashed, above right=0 and 0.3cm of codebooks.south east] (new_frame) {};
\node[mylittlesquare, fill=figc6, above right=\sepcodes and \sepcodes of new_frame.south west] (t60) {$Z_1^7$};
\node[mylittlesquare, fill=figc5, above=\sepcodes of t60] (t61) {$Z_2^6$};
\node[mylittlesquare, fill=figc4, above=\sepcodes of t61] (t62) {$Z_3^5$};
\node[mylittlesquare, fill=figc3, above=\sepcodes of t62] (t63) {$Z_4^4$};

\coordinate (pos_audio_embedding) at (2.75, 4.5);

\node[myrectangle3, rounded corners] (audio_embedding) at (pos_audio_embedding) {Audio\\Embedding};

\node[above=0.5cm of new_frame.north] (void_2) {};
\draw[-, thick] (quantizer.south) |- (void_2.center);
\draw[->, thick] (void_2.center) to (new_frame.north);
\draw[->, thick] (new_frame.south) |- (audio_embedding.west);

\node[myrectangle3, rounded corners, below=\sepblockslm of audio_embedding] (text_embedding) {Text\\Embedding};

\node[mytrapeze, align=center, below left=\sepblocks and \sepblocks of text_embedding, shape border rotate=270] (text_encoder) {Text\\Encoder};
\node[align=center, left=\sepblockslm of text_encoder] (text) {\texttt{Somber jazz}\\\texttt{track with}\\\texttt{piano melody}};

\draw[-, dashed] (text) to (text_encoder);
\draw[->, thick] (text_encoder) -| (text_embedding);

\node[right=0.9cm of audio_embedding] (lm_branch_block1) {};
\node[myrectangle3, rounded corners, right=1.2cm of audio_embedding, minimum width=2.5cm] (lm_sa_block1) {Causal\\Self-Attention};

\node[myrectangle3, rounded corners, figblue, fill opacity=0.2, minimum width=6.6cm, minimum height=4.4cm, draw=black, above left=-3.45cm and -6.2cm of lm_branch_block1] (lm) {};
\node[myrectangle3, rounded corners, dashed, draw=figblue, fill opacity=0.0, minimum width=5.2cm, minimum height=4.05cm, draw=black, above right=\sepblockslm and \sepblockslm of lm.south west] (lm_block1) {};

\node[myrectangle3, rounded corners, right=1.2cm of audio_embedding, minimum width=2.5cm] (lm_sa_block1) {Causal\\Self-Attention};
\node[myrectangle3, rounded corners, below=\sepblockslm of lm_sa_block1, minimum width=2.5cm] (lm_ca_block1) {Cross-Attention};
\node[myrectangle3, rounded corners, below=\sepblockslm of lm_ca_block1, minimum width=1.2cm] (lm_linear_block1) {Linear};
\node[sum, right=\sepblockslm of lm_ca_block1] (lm_sum_block1) {};
\node[myrectangle3, rounded corners, right=\sepblockslm of lm_sum_block1, minimum width=1.2cm] (lm_ln_block1) {Layer\\Norm};

\draw[-] (audio_embedding) to (lm_sa_block1);
\draw[-] (lm_sa_block1) to (lm_ca_block1.north);
\draw[-] (text_embedding) to (lm_ca_block1.west);
\draw[-] (lm_ca_block1) to (lm_linear_block1);
\draw[-] (lm_linear_block1) -| (lm_sum_block1.south);

\node[above=0.1cm of lm_sa_block1.north] (void_block1) {};
\draw[-] (lm_branch_block1.center) |- (void_block1.east); 
\draw[-] (void_block1.east) -| (lm_sum_block1.north);
\draw[-] (lm_sum_block1) to (lm_ln_block1.west);

\node[myrectangle3, rounded corners, dashed, draw=figblue, fill opacity=0.0, minimum width=0.75cm, minimum height=4.05cm, draw=black, right=0.2cm of lm_block1] (lm_blocklast) {\texttt{\dots}};
\node[below=0.cm of lm_blocklast.center] (dots) {\texttt{\dots}};

\node[myrectangle3, rounded corners, right=0.3cm of dots, minimum width=1.6cm] (lm_out) {Token\\Prediction};

\node[above=1.2cm of lm_out.center] (l_ce) {$\textcolor{figdarkergreen}{\mathcal{L}_\mathrm{CE}}$};
\draw[-, thick, loosely dotted, figdarkergreen] (lm_out) to (l_ce);

\node[above left=0cm and 0cm of lm_block1.south east] () {\texttt{$\times L$}};

\node[myrectangle3, minimum width=20pt, minimum height=2.67cm, dashed, right=0.3cm of lm_out] (new_frame_out) {};
\node[mylittlesquare, fill=figc7, above right=\sepcodes and \sepcodes of new_frame_out.south west] (t70) {$Z_1^8$};
\node[mylittlesquare, fill=figc6, above=\sepcodes of t70] (t71) {$Z_2^7$};
\node[mylittlesquare, fill=figc5, above=\sepcodes of t71] (t72) {$Z_3^6$};
\node[mylittlesquare, fill=figc4, above=\sepcodes of t72] (t73) {$Z_4^5$};

\draw[-] (dots.east) to (lm_out); 
\draw[->, thick] (lm_out) to (new_frame_out.west);

\node[above=0.4cm of new_frame_out] (void_frame) {};
\draw[-, dashed, thick] (new_frame_out.north) to (void_frame.center);
\draw[->, dashed, thick] (void_frame.center) to (quantizer.south east);

\end{tikzpicture}
}

    \caption{MusicGen framework. The EnCodec audio auto-encoder (top) encodes the waveform and audio tokens (middle) are obtained by discretizing the encoded audio with the RVQ multi-stage quantizer. The resulting audio tokens are then passed along text embeddings (bottom-left) to a Transformer-style language model with $L$ layers (bottom-right). The language model auto-regressively estimates the next token (right) according to the "delay" decoding strategy \cite{kharitonov2022textfree}.
    At the time step $t=7$, our proposed method MusicGen-MMD regularizes the EnCodec bottleneck with the loss \textcolor{figc0}{$\mathcal{L}_\mathrm{inde}$}, thereby promoting independence between the delayed codes $\{ Z_1^7, Z_2^6, Z_3^5, Z_4^4 \}$ produced by RVQ.
    }
    \label{fig:diagram}
\end{figure*}
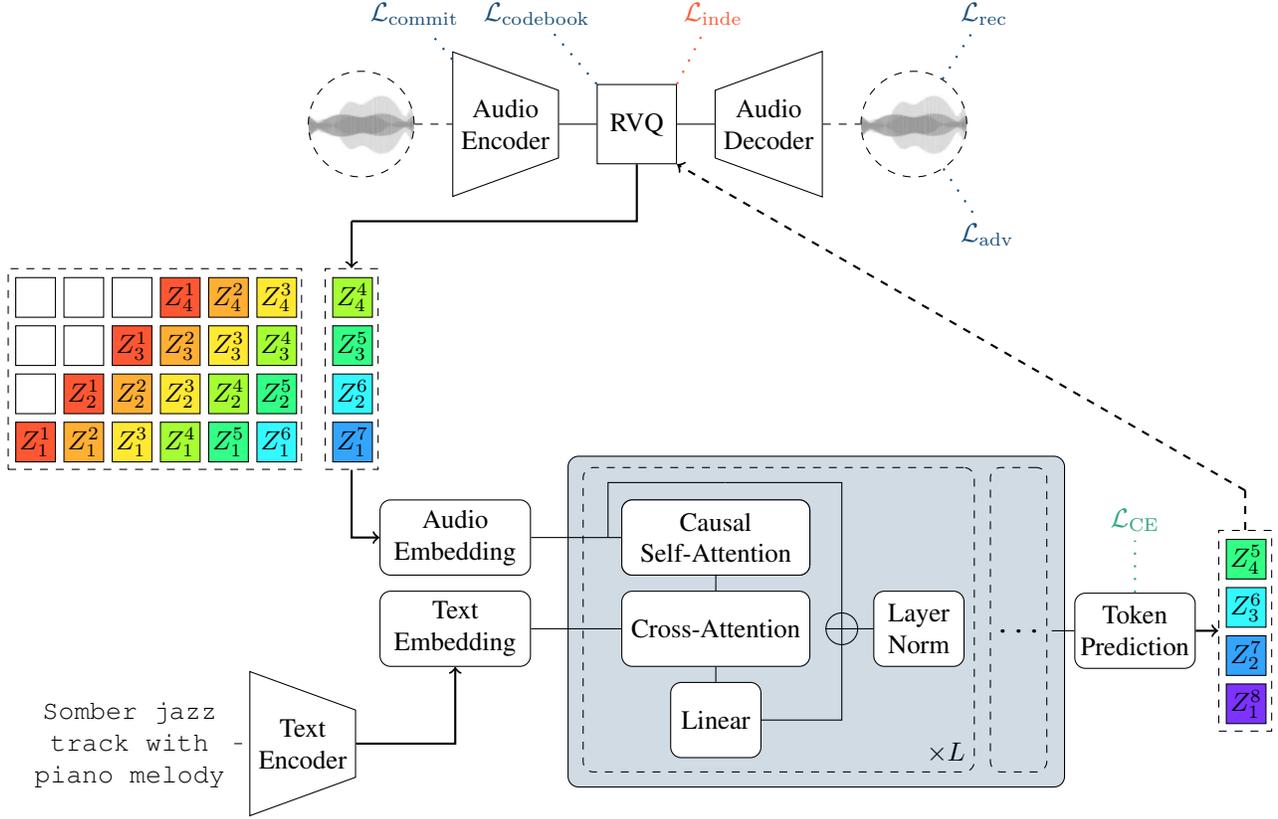

%% file: figures/algo.tex
\definecolor{figc0}{HTML}{ff5733}

\begin{algorithm}[tb]
   \caption{MMD Optimization}
   \label{alg:training}
\begin{algorithmic}
\STATE {\bfseries Input:} Training macro-batch $X$ \textcolor{blue}{\texttt{\%B,L}}
\STATE Encode $X_e = \mathcal{E}_\theta(X)$ \textcolor{blue}{\texttt{\%B,T,D}}%
\STATE Quantize $Z = \mathcal{Q}(X_e)$ \textcolor{blue}{\texttt{\%B,K,T,N}} %
\STATE \textit{Optional: Apply ``delay''
$Z_{.,k}^{(t)} = Z_{.,k}^{(t-k+1)}$}
\STATE Group time with batch axes $Z_{.,k} \leftarrow Z_{.,k,.}$ \textcolor{blue}{\texttt{\%B*T,K,N}}
\FOR{codebook index $k \in \{ 1, \dots, K \} $}
    \STATE Sample permutation $\pi \sim \mathcal{U}(\mathcal{S}_{BT})$
    \STATE Shuffle batch axis
    $\{ \bar{Z}_{i,k} \}_{i=1}^{BT} = \{ Z_{\pi(i), k} \}_{i=1}^{BT} $
\ENDFOR
\STATE Compute independence loss \eqref{eq:mmd_estimator} \textcolor{figc0}{$\mathcal{L}_\mathrm{inde}$}$ = \mathrm{MMD}(\mathbb{P}_Z || \mathbb{P}_{\bar{Z}})$
\end{algorithmic}
\end{algorithm}

%% file: sections/xp.tex
\input{figures/figure_mmd}

\definecolor{figblue}{HTML}{154c79}
\definecolor{figdarkergreen}{HTML}{2ba481}

\subsection{Models and Hyperparameters}

\textbf{Auto-encoder}: We use the 32kHz configuration of EnCodec \cite{defossez2023high} as our audio tokenizer.
EnCodec is a convolutional encoder-decoder model producing embeddings at 50 Hz for input waveforms sampled at 32 kHz. Each embedding is modeled by a RVQ scheme using 4 codebooks with $2^{11}=2048$ entries each, which leads to an effective bitrate of 2.2kB.${\mathrm{s}^{-1}}$. 
The model is trained with a reconstruction loss (\textcolor{figblue}{$\mathcal{L}_\mathrm{rec}$}) using a combination of $L^1$ and $L^2$ losses on the mel-spectrogram using multiple time resolutions (MSSpec), and a $L^1$ loss on the time signal. A multi-scale STFT discriminator is used to increase the reconstruction quality through adversarial training (\textcolor{figblue}{$\mathcal{L}_\mathrm{adv}$}), and a feature matching loss is added for the training of the generator \cite{kumar2019melgan}.
The quantizer is trained with the codebook loss (\textcolor{figblue}{$\mathcal{L}_\mathrm{codebook}$}), and the encoder is additionally trained with a commitment loss pulling the encoder outputs closer to the learnt embeddings (\textcolor{figblue}{$\mathcal{L}_\mathrm{commit}$}).
Models are trained for 600k steps on 8 V$100$ GPUs with the Adam optimizer, using $\beta_1=0.5$, $\beta_2=0.9$, a learning rate of $3 \cdot 10^{-4}$, a batch size of $64$ and segments of $1$ second cropped at random in audio sequences.
\newpage

\textbf{Language Model}: We train the same  Transformer model as MusicGen-small \cite{copet2023musicgen}, consisting of several Transformer-style layers for a total number of 300M parameters. Each layer comprises a causal self-attention module, a module computing cross-attention between the current signal and the conditioning text representation, a fully-connected block with ReLU, and a residual connection skipping from the layer's input. Sinusoidal positional encoding is used to embed the current time step \cite{vaswani2023attention}.
The decoding strategy for all models is the "delay" pattern \cite{kharitonov2022textfree}.
The model is trained on cross-entropy (\textcolor{figdarkergreen}{$\mathcal{L}_\mathrm{CE}$}) for 1M steps on 32 V$100$ GPUs with the AdamW optimizer, using $\beta_1=0.9$, $\beta_2=0.95$, a batch size of $192$, and audio sequences of $30$ seconds. We use a cosine learning rate schedule with a $4000$-steps warmup. Exponential moving average with a decay of $0.99$ is used to recursively smooth model weights. Top-250 sampling is used with a temperature of $1$ during inference  \cite{fan-etal-2018-hierarchical}.
The EnCodec audio codec and the text encoder are frozen during the training of the language model.

\textbf{Text Conditioning}: We use the T5 Transformed-based text encoder \cite{raffel2023exploring}. 
Metadata such as key, tempo or instrumentation are concatenated to the text description. 
We implement classifier-free guidance when sampling from the model’s logits, as in \cite{kreuk2023audiogen}. Therefore, we drop the conditioning signal with a probability of $0.2$ during training, and at inference we use a guidance strength of $3.0$.

\definecolor{figc0}{HTML}{ff5733}

\textbf{Independence Loss}: We use a weight of $10^3$ for the independence loss \textcolor{figc0}{$\mathcal{L}_\mathrm{inde}$}, computed in a separate backward. All the other losses are optimized as in \citep{defossez2023high}.
We choose this value empirically by selecting the largest weighting factor that did not degrade the traditional EnCodec loss, as detailed in the ablation study in Section~\ref{sec:ablation-scale}.
The RKHS $\mathbb{H}$ is equipped with the multi-scale Gaussian kernel $k(x, y) = \sum_{\sigma_i} e^{- || x - y ||^2 / 2 \sigma^2_i}$ with radii $\sigma_i \in \{ 0.1, 1, 5, 10, 20, 50 \} $. Therefore, it 
satisfies $\MMDH(\joint || \factor) = 0 \iff \joint = \factor$ (see Section~\ref{sec:mutual_information}). 
We let the kernel functions fixed throughout training, although optimizing the standard deviations $\sigma$ could lead to a better lower-bound of the true $\MMD$ in \eqref{eq:mmd_kernel}. This is because the distributions $\joint$ and $\factor$ are being learnt as we compute the $\MMDH$ estimator, therefore measuring the optimality of the chosen kernel $k(\cdot, \cdot)$ (or equivalently RKHS $\mathbb{H}$) is intrinsically hard. Furthermore, this would require a significant amount of energy spent in extensive grid searches, which we believe was not the focus of this study.
We further justify the choice of the multi-scale Gaussian kernel in Section~\ref{sec:kernel}.

Unless mentioned otherwise, we use the decoding strategy adaptation proposed in Section~\ref{sec:method} for the "delay" pattern \cite{kharitonov2022textfree}.
We noticed in our experiments that although the estimator \eqref{eq:mmd_estimator} is unbiased, a high batch size is required to reduce the variance of the estimator and properly optimize the objective \textcolor{figc0}{$\mathcal{L}_\mathrm{inde}$}.
We maximize the macro-batch size $B$ by accumulating $32$ batches, which results in $B=\mathrm{batches} \times \mathrm{batch} \mathrm{size} \times \tilde{T} / \mathrm{gpus}  = 1280$ samples per GPU).
We make these samples fit on a V$100$ GPU by using gradient checkpointing during encoding to compute the independence loss in a separate computational graph, which significantly reduces the amount of GPU memory used, at a minor increase in training time.

\subsection{Datasets}

We use 20K hours of licensed music to train both EnCodec and the language model. The training dataset is composed of an internal dataset of 10K high-quality music tracks, and the ShutterStock and Pond5 music data
collections\footnote{www.shutterstock.com/music www.pond5.com}, respectively consisting of 25K and 365K music tracks. 
All datasets comprise full-length music samples recorded at 32 kHz, accompanied by metadata including a textual description and supplementary details such as genre, key, tempo, etc. 
For comparison of the proposed method to the baselines, we employ the MusicCaps benchmark \cite{agostinelli2023musiclm} as our primary evaluation dataset. MusicCaps comprises 5.5K samples, each lasting ten seconds and curated by expert musicians. We resample all samples to 16kHz for fairness.
For ablation studies, we rely on a held-out internal evaluation set featuring 528 music tracks.

\input{tables/table_music}

\input{tables/table_mmd}

\subsection{Evaluation Metrics}
We conduct a comprehensive evaluation using both objective and subjective metrics. Objective functions include the Fréchet Audio Distance (FAD) \cite{kilgour2019frechet} computed as the distance between Gaussian distributions fitted on DNN-obtained embeddings of the real and generated samples.
As highlighted in \cite{gui2024adapting}, using FAD can lead to wrong interpretations if using irrelevant embeddings. We therefore use various embeddings such as CLAP-Laion
(contrastive learning audio pretraining), MERT-4 (acoustic music understanding) and VGGish (audio feature classification)\footnote{We compute all these scores using the official repository https://github.com/microsoft/fadtk associated to \cite{gui2024adapting}.}.
To complement this, akin to \cite{yang2023diffsound}, we calculate the KL-Divergence between the outputs of the Patch-Out-Transformer\footnote{https://github.com/kkoutini/PaSST} audio classifier \cite{Koutini_2022}, utilizing the original and generated audio as inputs. 
These metrics deliver insights into complementary aspects of the generated audio, namely quality, fidelity and high-level semantics.

For subjective evaluation, we conducted a MUSHRA-style \ac{mos} test, where 11 annotators were each asked to rate 12 samples each with a single number between 0 and 100 representing the overall music quality, including audio quality as well as consistency and likelihood of the harmonic, melodic and rhythmic structure. The ground-truth reference was given (and hidden among the samples for rating) as an anchor representing a music track with maximum music quality. The files rated by the annotators were randomly drawn from the MusicCaps dataset, normalized at -14dB LUFS\cite{itu}. The text description was not shown during the test. 
See Appendix~\ref{appendix:mushra} for more details.
We also run a second subjective evaluation with annotators recruited via Amazon Mechanical Turk: results and methodology are reported in Appendix~\ref{appendix:mos}.

\subsection{Baselines}

We compare our proposed method trained for music generation to the original MusicGen model without independence loss \cite{copet2023musicgen}, as well as other state-of-the-art latent diffusion baselines such as the text-to-music version of AudioLDM2 \cite{liu2023audioldm2}\footnote{
https://github.com/haoheliu/AudioLDM2} (denoted as AudioLDM2-Music in the following) 
, its predecessor AudioLDM \cite{liu2023audioldm}\footnote{
https://github.com/haoheliu/AudioLDM}, 
and Mustango \cite{melechovsky2023mustango}\footnote{https://github.com/AMAAI-Lab/mustango}. 
For completeness we also include other language modelling baselines such as MusicLM \cite{agostinelli2023musiclm}, Noise2Music \cite{huang2023noise2music} and the recent audio fondational model UniAudio \cite{yang2023uniaudio}. For these however, we were not able to evaluate these baselines as the public implementation was not made available for the given text-to-music generation task, and therefore reported results from the original papers directly.

\input{tables/table_kernels}

%% file: figures/figure_mmd.tex
\definecolor{figc0}{HTML}{ff5733}

\newcommand{\w}{0.35\textwidth}
\newcommand{\h}{0.28\textwidth}
\newcommand{\xs}{0.\textwidth}
\newcommand{\ys}{0.\textwidth}

\pgfplotsset{
    every axis plot/.append style={line width=2pt},
    discard if not/.style 2 args={
        x filter/.append code={
            \edef\tempa{\thisrow{#1}}
            \edef\tempb{#2}
            \ifx\tempa\tempb
            \else
                \def\pgfmathresult{inf}
            \fi
        }
    }
}

\begin{figure*}

\scalebox{0.97}{
\begin{tikzpicture}
  \begin{axis}[
    xlabel={MMD Weighting Factor},
    ylabel={MMD},
    xtick=data,
    xticklabels={$0$,$10^0$,$10^1$,$10^2$,$10^3$,$10^4$},
    name=mmd,
    width=\w,
    height=\h,
    xmode=log,
  ]
    \addplot+[blue, line width=0.1mm, mark options={fill=blue}] table[x=scale, y=mmd, col sep=comma]{scales_grid.csv};
  \end{axis}

  \begin{axis}[
    xlabel={MMD Weighting Factor},
    ylabel={$\mathcal{I} (\%)$},
    xtick=data,
    xticklabels={$0$,$10^0$,$10^1$,$10^2$,$10^3$,$10^4$},
    name=mi,
    width=\w,
    height=\h,
    at={(mmd.south east)},
    xshift=0.075\textwidth,
    ylabel style={yshift=-0.15cm},
    xmode=log,
  ]
    \addplot+[orange, line width=0.1mm, mark options={fill=orange}] table[x=scale, y=mi, col sep=comma]{scales_grid.csv};

  \end{axis}

  \begin{axis}[
    xlabel={MMD Weighting Factor},
    ylabel={MSSpec},
    xtick=data,
    ytick={0.106, 0.108, 0.110},
    yticklabels={0.106, 0.108, 0.110},
    xticklabels={$0$,$10^0$,$10^1$,$10^2$,$10^3$,$10^4$},
    name=mi,
    width=\w,
    height=\h,
    at={(mi.south east)},
    ylabel style={yshift=-0.1cm},
    xshift=0.095\textwidth,
    xmode=log,
  ]
    \addplot+[black, line width=0.1mm, mark options={fill=black}] table[x=scale, y=msspec, col sep=comma, mark color=black]{scales_grid.csv};

  \end{axis}
  
\end{tikzpicture}
}
    \caption{MMD, total correlation of EnCodec codes and MSSpec loss computed on our internal set. 
    MSSpec is the combination of $L1$ and $L2$ losses on the multi-resolution mel-spectrogram, used for reconstruction in EnCodec.
    The horizontal axis shows the weighting factor used for the MMD loss \textcolor{figc0}{$\mathcal{L}_\mathrm{inde}$}. The total correlation $\mathcal{I}$ is computed on the whole 250k-samples training set for minimal bias in the histogram approximation. It is computed between two codebooks taken at random, averaged over five codebook couples, and expressed as a ratio to the entropy of the joint distribution (in \%).}
    \label{fig:scales}
\end{figure*}
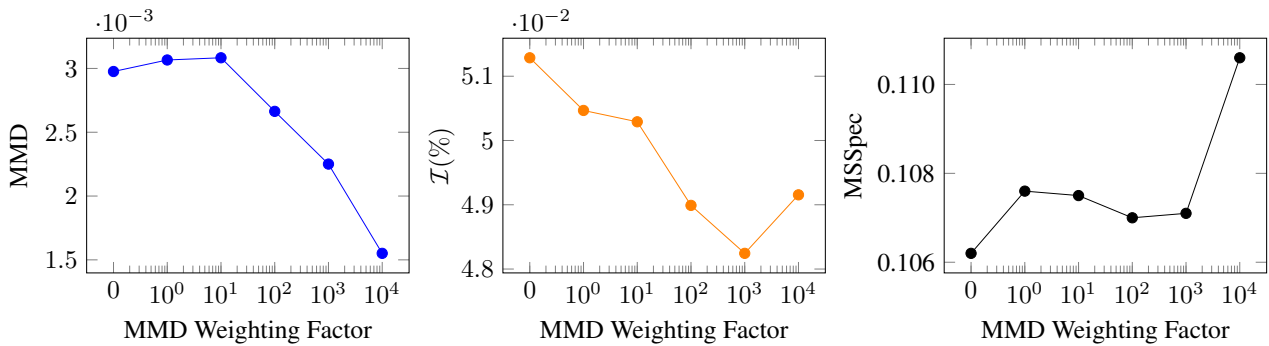

%% file: tables/table_music.tex
\begin{table*}[h]
    \centering

    \caption{Text-to-music generation on MusicCaps. Asterisks$^{\ast}$ mean that we report figures from the related papers as the public implementation was not available for the given text-to-music generation task. Mustango was trained on an augmented version of MusicCaps, therefore we put it aside the other baselines. For the subjective metric (OVRL), mean and 95 \% confidence intervals are showed. The samples presented were 10 second long sampled at 16kHz, which matches the training conditions of Mustango, AudioLDM and AudioLDM2-Music. In comparison MusicGen and MusicGen-MMD were trained on 30 second-long segments sampled at 32kHz.
     \vspace{0.5em}
    }
    \scalebox{0.85}{
    \begin{tabular}{l c|ccc|ccc}
    \toprule
         Model & \# params & 
         FAD$_{clap\mathrm{-}laion} \downarrow$ & 
         FAD$_{MERT\mathrm{-}4} \downarrow$ & 
         FAD$_{vgg} \downarrow$ & 
         KL $\downarrow$ &
         CLAP (\%) $\uparrow$ & 
         OVRL. $\uparrow$ \\  \hline
         
    \textit{Ground-Truth} & & - & - & - & - &
    \textit{38} & \textit{97.95 $\pm$ 1.13} \\ \hline
    
    \textit{Mustango} & 1.4 B & \textit{0.07} & \textit{1.65} & 
    \textit{1.56} 
    & \textit{0.71} & \textit{37} & \textit{49.26 $\pm$ 4.21} \\ \hline

    MusicLM$^{\ast}$ & 860 M & - & - & 4.0 & - & - & - \\
    Noise2Music$^{\ast}$ & 1.3 B & - & - & 2.1 & - & - & - \\
    UniAudio$^{\ast}$ & 1 B & - & - & 3.65 & 1.87 & - & -\\ \hline
    
    AudioLDM & 416 M & 0.18 & 4.18 & 
    3.52
    & 1.42 & \textbf{35} & 56.29 $\pm$ 4.35 \\
    
    AudioLDM2-Music & 347 M & 0.25 & 4.30 & 
    4.71 
    & 1.31 & 31 & 69.43 $\pm$ 3.42 \\
    
    MusicGen & 300 M & 0.16 & 1.57 &  
    3.60 
    & 1.22 & 31 & 62.54 $\pm$ 3.68 \\
    
    MusicGen-MMD (ours) & 300 M & \textbf{0.14} & \textbf{1.45} & 
    \textbf{2.98} 
    & \textbf{1.18} & 32 & \textbf{74.75 $\pm$ 3.68} \\
    
     \hline \bottomrule
    \end{tabular}}
     \vspace{-0.5em}
    \label{tab:musiccaps}
\end{table*}

%% file: tables/table_mmd.tex
\begin{table}[h]
    \caption{Text-to-music generation results on held-out test set. All models have 300M parameters.
     \vspace{0.5em}
    }
    \hspace{-0.15cm}
    \scalebox{0.87}{
    \begin{tabular}{l ccc}
    \toprule
    MusicGen Configuration & FAD$_{vgg} \downarrow$ & KL $\downarrow$ & CLAP (\%) $\uparrow$ \\ \hline
    \textit{Ground-truth} & - & - & \textit{38} \\ \hline
    Delay \cite{copet2023musicgen} & 0.95 & \textbf{0.45} & 37\\
    Delay w/ MMD-Parallel & 0.90 & \textbf{0.45} & 37\\
    Delay w/ MMD (proposed) & \textbf{0.59} & 0.46 & 37\\
    Flatten & 0.69 & 0.46 & \textbf{39}\\
     \hline \bottomrule
    \end{tabular}
    }
     \vspace{-0.5em}
    \label{tab:ablation-pattern}
\end{table} 

%% file: tables/table_kernels.tex
\begin{table}[h]
    \caption{MMD, total correlation and reconstruction losses of EnCodec-MMD with various kernels evaluated on our internal dataset. We used a weight of 1000 for the MMD loss, and adapted the weighting factors of the MMD loss so that the magnitudes of the losses stayed approximately consistent across kernels.
     The total correlation $\mathcal{I}$ is computed on the whole 250k-samples training set for minimal bias in the histogram approximation. It is calculated between two codebooks taken at random, averaged over five codebook couples, and expressed as a ratio to the entropy of the joint distribution (in \%).
     \vspace{0.5em}
    }
    \centering
    \scalebox{0.87}{
    \begin{tabular}{l cc}
    \toprule
         Method & $\mathcal{I}$ (\%) $\downarrow$ & MSMelSpec $\downarrow$ \\ \toprule \hline 
         Multi-Scale Gaussian & 4.8 $\cdot 10^{-2}$ & \textbf{0.107} \\ 
         Squared Inverse & \textbf{4.1 $\cdot 10^{-2}$} & 0.127 \\
         Linear & 5.0 $\cdot 10^{-2}$ & 0.114 \\
         Quadratic & 4.9 $\cdot 10^{-2}$ & 0.118 \\ 
         \bottomrule
    \end{tabular}
    }
     \vspace{-0.5em}
    \label{tab:kernels}
\end{table}

%% file: sections/results.tex
We introduce our results section by running an analysis of the proposed independence-proxy loss with respect to the weighting factor used for optimization, and investigate its correlation with total correlation of the codes.
We follow by reporting objective and subjective metrics for music generation on the standard MusicCaps benchmark. 
Then, we proceed with an ablation study to show the efficiency of integrating the decoding strategy for MMD loss optimization.
We also test the generalizibility of our method by applying it to a different state-of-the-art audio codec, namely RVQGAN \cite{kumar2024highfidelity}, and we analyse the resulting performance in appendix \ref{appendix:dac}.
Finally, we conduct ablation studies with respect to other quantization schemes: results are reported in appendices \ref{appendix:dac},\ref{appendix:quantizers} and \ref{appendix:hierarchy}.

\subsection{MMD as an Independence-promoting Loss}
\label{sec:ablation-scale}

We show in Figure~\ref{fig:scales} the MMD, total correlation and MSSpec loss values for EnCodec codes (which are later used as tokens in our language model).
We show our grid search with respect to the scaling factor for the MMD loss.
We use our whole 250k-samples internal set for minimal bias in histogram approximation. The total correlation $\mathcal{I}$ is computed between two codebooks taken at random, averaged over five codebook couples, and expressed as a ratio to the entropy of the joint distribution (in \%).
We first observe that MMD overall correlates with the total correlation, which shows that our proposed loss is a reasonable independence proxy. 
Except for the large weighting factor of $10^4$, the MMD loss and total correlation diminish monotonously with respect to the weighting factor used for optimization, which qualifies the proposed criterion as a valid objective loss.
The MSSpec reconstruction loss remains unaffected except when using a very large scaling factor of $10^{4}$, for which the training seems perturbed, and where the total correlation does not seem to correlate with MMD anymore. We choose a factor of $10^3$ as it allows a maximal total correlation reduction without hurting the reconstruction loss. 

We show in Appendix~\ref{appendix:dac} that our method is generalizable to other codecs, by applying MMD optimization to the latent space of RVQGAN \cite{kumar2024highfidelity}, which is a state-of-the-art audio codec based on EnCodec. Our results support that MMD optimization can also be used to promote the independence of RVQGAN codes, in a similar fashion to what have demonstrated here for EnCodec codes.

\subsection{Text-to-Music Generation Benchmark}

We show objective and subjective evaluation results for music generation on MusicCaps in Table~\ref{tab:musiccaps}.
We observe that the objective metrics of Mustango are quite strong, as the model was trained on an augmented version of MusicCaps.
Our method MusicGen-MMD improves objective metrics over our own baseline MusicGen, and obtains better objective metrics than AudioLDM, AudioLDM2-Music, MusicLM and UniAudio. Noise2Music still obtains a better FAD result, although with a much larger architecture (1.3 B). Furthermore, we could not reproduce the results nor run other metrics (such as FAD with other embeddings) as the implementation was not made publicly available.
The subjective metric OVRL. obtained via the MUSHRA-style test indicates that our model MusicGen-MMD obtains the best performance, closely followed by AudioLDM2-Music. Then follow MusicGen, AudioLDM and finally Mustango.

\subsection{Decoding Strategy Matching}

We present the effect of integrating the language model decoding strategy to the MMD loss optimization. We train three models with the same language modeling configuration and the "delay" decoding stategy, but distinct EnCodec configurations: our baseline without MMD optimization (Delay), our proposed model  using the "delay" decoding strategy for optimizing the MMD (Delay w/ MMD) and our proposed model where the MMD optimizes does not integrate the decoding strategy (Delay w/ MMD-Parallel). 
Finally we train a MusicGen model using the "flatten" decoding strategy where the codebooks are flattened such that a single code is predicted at each time step. This effectively models the joint distribution $\mathbb{P}_Z$ instead of the factorized distribution $\mathbb{P}_{\bar{Z}}$.
Results are computed on our held-out test set and reported in Table~\ref{tab:ablation-pattern}.
Objective scores show that adapting the MMD optimization to the language modelling decoding strategy improves audio quality and fidelity, as our proposed method obtains a better FAD$_{vgg}$ than the one where the MMD criterion is not adapted to the language model decoding strategy. Our method even outperforms the MusicGen with "flatten" strategy on the FAD$_{vgg}$ score, which indicates that training the language model to predict the joint distribution by flattening the codebooks does not yield optimal performance, which we posit is due to increased training difficulty. In addition, the original frame rate of EnCodec is preserved, whereas MusicGen with "flatten" decoding largely increases the inference time, by a factor equal to the number of codebooks $K$.

\subsection{Kernel Function Ablation}
\label{sec:kernel}

We justify here how the choice of kernel function $k(\cdot, \cdot)$ impacts the reconstruction error of EnCodec and the total correlation of the codes.

First, the Gaussian kernel is a natural candidate as it is widely used in statistics and machine learning. Furthermore, we observed experimentally that using several standard deviations $\sigma_i$ increases the numerical robustness of the MMD computation, as unadapted values might make the exponentials in the Gaussian kernel collapse to values where the numerical rounding errors degrade the estimation of the MMD.
Using several $\sigma_i$ therefore enables us to avoid this pitfall, as we can expect at least some values to produce reliable estimates.  

We have conducted experiments with a variety of other kernels and provide the results in Table~\ref{tab:kernels}.
The squared inverse kernel is defined here as $k(x, y) = (1 + (|| x - y ||^2) / \sigma^2)^{-1}$ with $\sigma=12$, the linear kernel as $k(x, y) = x^T y$ and the quadratic kernel as $k(x, y) =  (x^T y)^2$.
We observe that the multi-scale Gaussian kernel achieves the most interesting trade-off, by obtaining the second lowest total correlation while outperforming all other kernel functions on reconstruction, thereby justifying its choice in our subsequent experiments.

%% file: sections/appendix.tex
\section{Proof of Kernel Formulation of MMD}
\label{appendix:proof}

This is the proof to \eqref{eq:mmd_kernel} and mostly uses material from \cite{mmd}.
First, the notion of feature mapping can be extended to the \textit{mean embedding} of a probability distribution \cite{mmd}. Given a probability distribution $\mathbb{P}_X$
we define its mean embedding $\mu_{\mathbb{P}_X} := \mathbb{E}_{X \sim \mathbb{P}_X} [\phi(X)] \in \mathbb{H}$ such that:
    \begin{equation}\label{eq:mean}
         \forall \, h \in \mathbb{H} : \,  \mathbb{E}_{X \sim \mathbb{P}_X} [h(X)] = \langle h, \mu_{\mathbb{P}_X} \rangle_{\mathbb{H}}
    \end{equation}
If $h$ is taken to be in a RKHS $\mathbb{H}$, the obtained MMD estimate is actually a lower-bound of the true MMD:
\begin{align*}
    \MMD(\mathbb{P}_Z || \mathbb{P}_{\bar{Z}}) &= \sup_{h, \norm{h} 
\leq 1} \mathbb{E}_{\mathbb{P}_Z}[h(\bar{Z})] - \mathbb{E}_{\mathbb{P}_{\bar{Z}}}[T(\bar{Z})] \\
& \geq \underbrace{\sup_{h \in \mathbb{H}, \norm{h} 
\leq 1} \mathbb{E}_{\mathbb{P}_Z}[h(\bar{Z})] - \mathbb{E}_{\mathbb{P}_{\bar{Z}}}[T(\bar{Z})]}_{\MMD_{\mathbb{H}}(\mathbb{P}_Z || \mathbb{P}_{\bar{Z}}) }.
\end{align*}
Using \eqref{eq:mean} in \eqref{eq:mmd} and the properties of $\mathbb{H}$, we can then compute the MMD between $\mathbb{P}_Z$ and $\mathbb{P}_{\bar{Z}}$ taking the supremum over the unit ball of $\mathbb{H}$ as:
\begin{align*}
    \MMD_{\mathbb{H}}(\mathbb{P}_Z || \mathbb{P}_{\bar{Z}}) &= \sup_{h \in \mathbb{H}, \norm{h} \leq 1} \mathbb{E}_{\mathbb{P}_Z}[h(\bar{Z})] - \mathbb{E}_{\mathbb{P}_{\bar{Z}}}[T(\bar{Z})] \\
    &= \sup_{h \in \mathbb{H}, \norm{h} \leq 1} \langle h, \mu_{\mathbb{P}_Z} - \mu_{\mathbb{P}_{\bar{Z}}} \rangle \\
    &= \norm{\mu_{\mathbb{P}_Z} - \mu_{\mathbb{P}_{\bar{Z}}}}_{\mathbb{H}} \\
    &= \langle \mu_{\mathbb{P}_Z}, \mu_{\mathbb{P}_Z} \rangle \footnotesize{- 2} \langle \mu_{\mathbb{P}_Z}, \mu_{\mathbb{P}_{\bar{Z}}} \rangle \footnotesize{+} \langle \mu_{\mathbb{P}_{\bar{Z}}}, \mu_{\mathbb{P}_{\bar{Z}}} \rangle,
\end{align*}
where we use the $1$-Lipschitz property of $h$ in the third line.
We can then use the definition of the mean embedding to obtain:
\begin{align*}
    \MMD_{\mathbb{H}}(\mathbb{P}_Z || \mathbb{P}_{\bar{Z}}) &=
     \, \, \hspace{0.2pt}
     \mathbb{E}_{Z_1 \sim \mathbb{P}_Z} \mathbb{E}_{Z_2 \sim \mathbb{P}_Z} \langle \phi(Z_1), \phi(Z_2) \rangle \\
    &+ \, \, \,
    \mathbb{E}_{\bar{Z}_1 \sim \mathbb{P}_{\bar{Z}}} \mathbb{E}_{\bar{Z}_2 \sim \mathbb{P}_{\bar{Z}}} \langle \phi(\bar{Z}_1), \phi(\bar{Z}_2) \rangle \\
    &- 2 \mathbb{E}_{Z_1 \sim \mathbb{P}_Z} \mathbb{E}_{\bar{Z}_2 \sim \mathbb{P}_{\bar{Z}}} \langle \phi(Z_1), \phi(\bar{Z}_2) \rangle.
\end{align*}
Finally, using the kernel definition in $\mathbb{H}$:
\begin{align*}
    \MMD_{\mathbb{H}}(\mathbb{P}_Z || \mathbb{P}_{\bar{Z}}) &= 
    \, \, \hspace{0.5pt}
    \mathbb{E}_{Z_1 \sim \mathbb{P}_Z} \mathbb{E}_{Z_2 \sim \mathbb{P}_Z} k(Z_1, Z_2) \\
    &+  \, \, \,
    \mathbb{E}_{\bar{Z}_1 \sim \mathbb{P}_{\bar{Z}}} \mathbb{E}_{\bar{Z}_2 \sim \mathbb{P}_{\bar{Z}}} k(\bar{Z}_1, \bar{Z}_2) \\
    &- 2 \mathbb{E}_{Z_1 \sim \mathbb{P}_Z} \mathbb{E}_{\bar{Z}_2 \sim \mathbb{P}_{\bar{Z}}} k(Z_1, \bar{Z}_2).
\end{align*}

\section{MMD Optimization on RVQGAN Codes}
\label{appendix:dac}

\input{figures/figure_dac_mmd}

We apply here our MMD optimization method on RVQGAN \cite{kumar2024highfidelity}, a state-of-the-art codec based on EnCodec. RVQGAN improves upon EnCodec by using lower-dimensional embeddings in the RVQ codebooks, thereby increasing codebook utilization. The authors also propose a new multi-scale STFT discriminator and various other techniques to increase the quality at lower-bitrate regimes.
Our aim here is to demonstrate that our independence-promoting criterion based on MMD optimization is generalizable to other codecs.
We employ the same setup as in our main experiments, and simply use RVQGAN in place of EnCodec, keeping the number of codebooks and the total bandwidth identical. 
We show the MMD loss, mutual information of RVQGAN codes and reconstruction losses in Figure~\ref{fig:dac-mmd}. We observe the similar trend compared to our method applied to EnCodec, with an even stronger correlation between the scale of the MMD loss and the mutual information, which implies that MMD optimization of the RVQGAN latent space also correlates with a more independence of the RVQGAN codes.

\section{MMD Optimization with Different Quantization Schemes}
\label{appendix:quantizers}

Product vector quantization (PVQ) is another multistage quantization method, where the input vector dimensions are split across $C$ groups and each group of dimensions is encoded by a codebook with dimensionality $N/C$. Although this scheme is typically non-hierarchical, since no priority is given to any particular codebook, a hierarchy can be introduced through hierarchical
dropout (PVQ-dropout). This means sampling a natural number $k \sim \mathcal{U}( \{ 1, \dots, K\} )$ and using only the \textit{first} $k$ codebooks for encoding (and putting the other codes to 0 before decoding). This quantizer dropout technique is also used in the RVQ-based SoundStream codec \cite{zeghidour2021soundstream}, however with a different intent: it allows the resulting codec to function at various bitrates without further adaptation at training time.

We employ here a similar setup as in Section~\ref{sec:ablation-scale}. We show in Table~\ref{tab:mmd_vs_mi} the MMD and total correlation values for EnCodec codes (which are later used as tokens in our language model), with the chosen scale factor of $10^{3}$.
We use our whole 250k-samples internal set for minimal bias in histogram approximation.
The total correlation $\mathcal{I}$ is computed between two codebooks taken at random, averaged over five codebook couples, and expressed as a ratio to the entropy of the joint distribution (in \%).
We observe that residual quantization introduces more dependence between codes compared to product quantization, although both induce a hierarchical structure in the codes space, which accounts for their high coding efficiency.
We also observe that our proposed MMD loss is able to curb both the MMD and total correlation of the PVQ w/ dropout codes, highlighting its versitality.

\begin{table}[h]
    \caption{MMD and total correlation of EnCodec codes. Results computed on complete 250k-samples internal set.
     \vspace{0.5em}
    }
    \centering
    \begin{tabular}{l cc}
    \toprule
         EnCodec Quantizer & MMD$\downarrow$ & $\mathcal{I}$ (\%) $\downarrow$ \\ \toprule \hline 
         RVQ & 9.9 $\cdot 10^{-4}$ & 5.1 $\cdot 10^{-2}$ \\ 
         RVQ w/ MMD & 9.9 $\cdot 10^{-5}$ & 4.8 $\cdot 10^{-2}$ \\ \hline
         PVQ w/ dropout & 3.7 $\cdot 10^{-5}$ & 3.8 $\cdot 10^{-2}$ \\ 
         PVQ w/ dropout + MMD & 4.5 $\cdot 10^{-7}$ & 3.0 $\cdot 10^{-2}$ \\ 
         \bottomrule
    \end{tabular}
     \vspace{-0.5em}
    \label{tab:mmd_vs_mi}
\end{table}

\section{Effect of Hierarchy in Quantized Audio Space}
\label{appendix:hierarchy}

We investigate here the performance of language models as a function of the quantization scheme used. 
We use three different quantizers for EnCodec: RVQ, which is our default quantizer, PVQ and PVQ-dropout. As explained in \ref{appendix:quantizers}, introducing a codebook dropout mechanism in PVQ naturally induces a hierarchical structure, as EnCodec will more regularly rely on the first few codebooks to reconstruct the audio. 
By looking at the contributions of individual codebooks (not shown here), we can observe a similar hierarchical structure for PVQ-dropout and RVQ, and no hierarchy in PVQ codes.
We subsequently trained three language models with their respective EnCodec configurations (RVQ, PVQ, PVQ w/ dropout) and the same language model configuration. Objective results on our held-out test set are reported in Table~\ref{tab:ablation-hierarchy}. We observe that the model using PVQ has low objective scores, while that using PVQ w/ dropout obtains much better objective scores at language modeling, somewhat close yet still inferior to the RVQ-equipped model, which seems to be the best strategy here and demonstrates the high coding efficiency of residual vector quantization.
This seems to indicate that hierarchical structure in the token space leads to better language modeling performance, which we posit is due to the language models being able to rely on its first few codebooks in case its modeling capacity it too limited.
On the other hand, as we indicated in the main paper, promoting independence between codes for exact modeling of the codebook distributions is also theoretically motivated and experimentally demonstrated.
This means there is potentially a trade-off to seek between hierarchy and independence in the codes space. 
The first is obtained via structural properties of the used quantizer e.g. residual quantization or dropout, and the second can be tuned via independence optimization as proposed in this paper. We argue that the complimentary nature of these solutions allows for a control over this trade-off for optimal audio generation performance.

\begin{table}[h]
    \caption{Text-to-Music generation of MusicGen with various quantization schemes for EnCodec tokenizer. Results are shown on the held-out test set. All models have 300M parameters.
     \vspace{0.5em}
    }
    \centering
    \scalebox{0.89}{
    \begin{tabular}{l ccc}
    \toprule
   EnCodec Quantizer & FAD$_{vgg} \downarrow$ & KL $\downarrow$ & CLAP (\%) $\uparrow$\\ \hline
    RVQ & \textbf{0.97} & \textbf{0.45} & \textbf{37} \\
    PVQ w/ dropout & 1.26 & \textbf{0.45} & 36 \\
    PVQ & 1.66 & 0.49 & 36 \\
     \hline \bottomrule
    \end{tabular}
    }
     \vspace{-0.5em}
    \label{tab:ablation-hierarchy}
\end{table}

\section{Mutual Information of State-of-the-art Codecs}
\label{sec:mi-codecs}

\input{figures/figure_codecs_music}

\input{figures/figure_codecs_speech}

We provide here additional insights into various state-of-the-art speech and music codecs. For all these codecs, we compute the mutual information between \textit{individual} codebooks and all the remaining codebooks.

\textit{Music Codecs}

We include in Figure~\ref{fig:mi-fma-pop} the mutual information of codes computed on the public music dataset FMA-Pop proposed in [4], as we found out that MusicCaps did not provide enough samples for reliable joint density histogram computation. 
Our results seem to show that both the original EnCodec (EnCodec-24kHz, \cite{defossez2023high}) and the 4-level MusicGen variant of EnCodec (EnCodec-32kHz, \cite{copet2023musicgen}) suffer from relatively high inter-codebook dependence, and that indeed RVQGAN obtains a large decrease of mutual information between codebooks, which can arguably be attributed to the choice of lower codebook dimensionality as suggested by the authors \cite{kumar2024highfidelity}. 
However, this does not mean that there is no room for improvement on this basis, as the independence-
promoting mechanism for RVQGAN is structural, based on limitation of the amount of information learnable by a single codebook, and can also be completed with explicit MMD optimization, as we have demonstrated in Appendix~\ref{appendix:dac}.

\textit{Speech Codecs}

We compute the mutual information between the codebooks of SpeechTokenizer \cite{zhang2024speechtokenizer} and FACodec \cite{ju2024naturalspeech} on LibriSpeech using 32k 200-second-samples and show the results on Figure~\ref{fig:mi-speech}.
We compared to the results of the original EnCodec (EnCodec-24kHz, \cite{defossez2023high}) which was trained on audio data including speech). 

We observe that the mutual information between EnCodec and SpeechTokenizer codebooks and the other codebooks decrease monotonously with the codebook index, which is expected given the residual quantization scheme. For SpeechTokenizer we observe that the mutual information between the first codebook and the remaining codebooks is by far the largest across codebooks. Indeed, although the information in codebook 1 is specifically distilled from HuBert, there is actually no mechanism (unlike FACodec) that specifically prevents the codebooks 2:8 to use information from codebook 1. Yet, the authors confirm experimentally that the speaker-specific information is contained in the codebooks 2:8 and that codebook 1 contains mostly content information. This poses the question of how mutual information is exactly related to such semantics. For FACodec, the mutual information between the prosody stream and the content stream is also relatively high, but the mutual information between all other pairs of streams is very low, which shows some successful disentanglement. Overall it seems FACodec boasts the best level of disentanglement among the considered baselines.
However, one must mention that speech semantic are much easier to investigate via the use of explicit audio properties (F0, phoneme label, ...) as opposed to music semantics. This enables for instance FaCodec to use
gradient-reversal layers for supervising the disentanglement of their streams such as e.g. prosody and timbre. Our independence-promoting method, on the other hand, is fully unsupervised and domain-agnostic.

\section{MUSHRA-style MOS Listening Test}
\label{appendix:mushra}

\begin{figure}[h]
    \centering
    \includegraphics[width=0.48\textwidth]{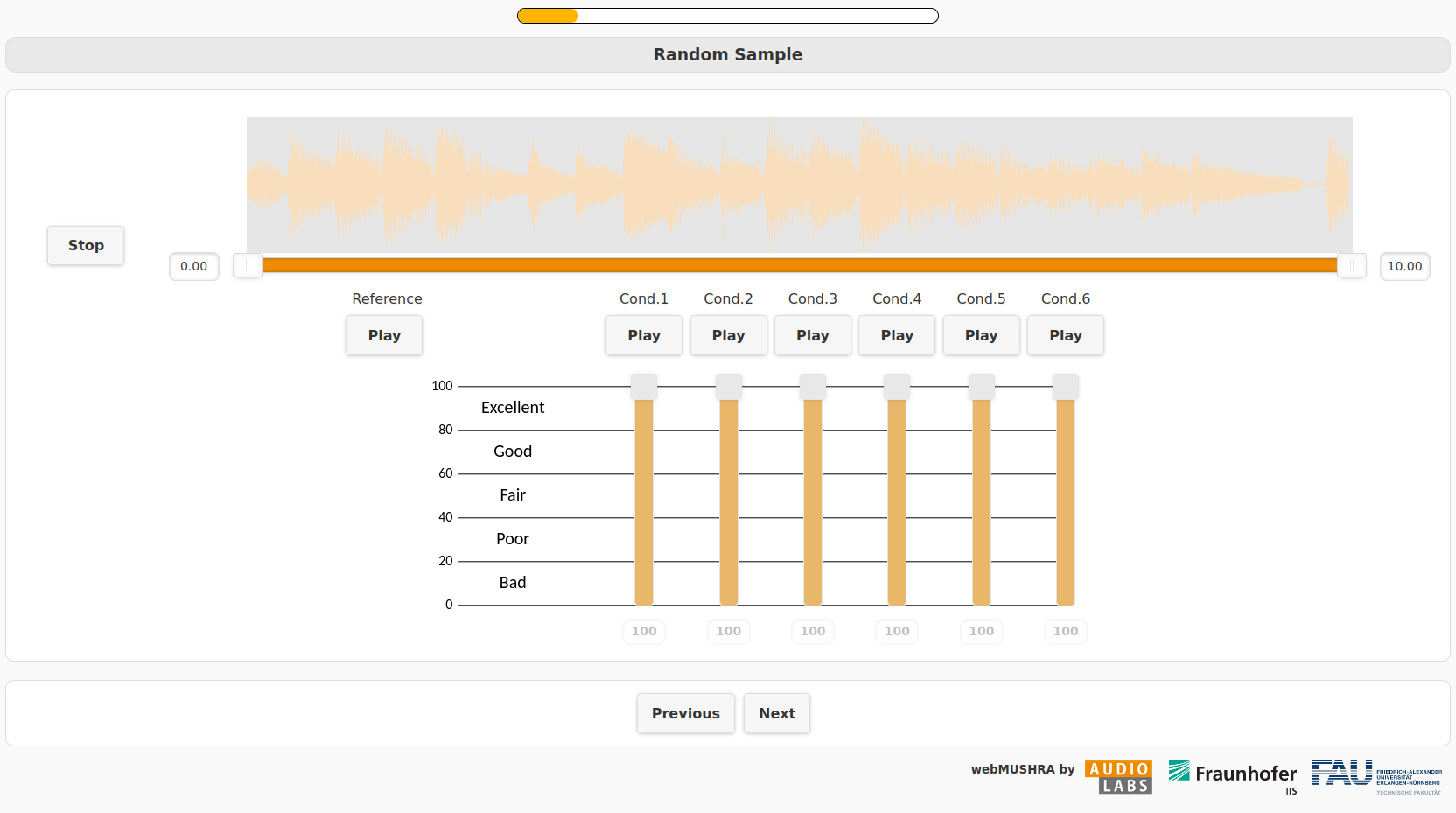}
    \caption{The MUSHRA listening test interface. Annotators listen to each sample and adjust the vertical bar on a continuous scale between 0 and 100. The reference track is given on the left and also hidden among the samples for rating.}
    \label{fig:interface-mushra}
\end{figure}

Our subjective benchmark is a MUSHRA-style MOS listening test produced with the $\texttt{webMUSHRA}$\footnote{https://github.com/audiolabs/webMUSHRA} tool with \texttt{pymushra}\footnote{https://github.com/nils-werner/pymushra} server management. In total, 12 annotators are asked to rate on a scale of 0 to 100 the overall quality of 12 10-second samples, whose descriptions were taken at random from the MusicCaps test set. All samples are normalized at -14dB LUFS\cite{itu}.
All annotators have a solid background either in audio or music processing.
The instructions given on the training page are as follows:
\textit{``You are asked here to rate the different samples provided with respect to the reference. The rating should reflect the overall quality, comprising music quality, harmonic, melodic and rhythmic structure. You are not asked to rate the distance of the samples with respect to the reference in terms of sound similarity but along the aforementioned dimensions (quality, structure, consistency).''}
The presentation order of the samples is randomized for each listener differently, and all 12 listeners listened to all of the samples.
A snapshot of the interface for a randomized trial is shown on Figure~\ref{fig:interface-mushra}. 
Inspired by the CrowdMOS guidelines, we excluded the annotations where reference track was rated below 85. We further excluded one annotator that systematically rated all generated samples below 50, resulting in the number of 11 annotators reported in the main paper.

\section{MOS Evaluation with Amazon Mechanical Turk}
\label{appendix:mos}

We conducted a second subjective evaluation using the same subjective benchmark as \cite{copet2023musicgen, kreuk2023audiogen}, inspired by \cite{yang2023diffsound}. Human raters are sollicited via the Amazon Mechanical Turk platform and receive compensation meeting the American minimum wage. They assess two primary aspects of the audio signal: (i) overall quality (OVRL.), rated as the perceptual quality on a scale of 1 to 100; (ii) relevance to the text input (REL.), rated as the alignment between the audio and the text prompt on a scale of 1 to 100. Subjects evaluate 100 randomly selected files from the MusicCaps and AudioCaps test set, for music generation and general audio generation respectively. Each sample is assessed by at least 5 raters. The CrowdMOS\footnote{http://www.crowdmos.org/download/} package is employed to filter out noisy annotations and outliers. This involves the exclusion of annotators who did not listen to the full recordings, those who rated the reference recordings below 85, and other CrowdMOS guidelines \cite{ribeiro2011crowdmos}.
Results are shown in Table~\ref{tab:music-results-turk}, and show that our method MusicGen-MMD is still ranking very high among baselines in terms of subjective ratings. However, the differences between the methods are rather marginal.
The main difference between the methodology of the two tests resides in the recruitment of subjects (which is specified by the MUSHRA ITU-R BS.1534-0 recommendation).
For the MUSHRA-style MOS experiment reported in the paper, we recruited confirmed audio listeners, and made sure that their setup was reliable (quiet environments, high-quality noise-canceling headphones...). 
On the other hand, we did not have any insight in the setups that subjects used in the MOS listening test in appendix. It is rather common than Mechanical Turk raters have low-quality setups, in potentially noisy environments, are not trained audio experts, and have little incitement for performance due to the low monetary retribution.
For this reason, we believe the MUSHRA-style MOS evaluation reported in Table~\ref{tab:musiccaps} is more reliable as the one conducted with Mechanical Turk raters, and therefore reported the first one in the main paper, and the second one in this appendix out of completeness.

\begin{table}[h]
    \caption{Subjective evaluation for text-to-music generation on MusicCaps. Mustango was trained on an augmented version of MusicCaps, therefore we put it aside the other baselines. Mean and 95 \% confidence intervals are showed. The samples presented were 10 second long sampled at 16kHz, which matches the training conditions of Mustango, AudioLDM and AudioLDM2-Music. In comparison MusicGen and MusicGen-MMD were trained on 30 second-long segments sampled at 32kHz.
     \vspace{0.5em}
    }
    \centering
    \scalebox{0.8}{
    \begin{tabular}{l c|cc}
    \toprule
         Model & \# params & OVRL. $\uparrow$ & REL. $\uparrow$ \\ \hline

    \textit{Ground-Truth} & - & \textit{92.49 $\pm$ 1.65} & \textit{92.89 $\pm$ 1.38} \\ \hline
    \textit{Mustango} & 1.4 B & \textit{81.24 $\pm$ 2.43} & \textit{84.27 $\pm$ 1.95} \\ \hline
    AudioLDM & 416 M & \textbf{84.70 $\pm$ 2.25} & 84.20 $\pm$ 3.12 \\
    AudioLDM2-Music & 347 M & 81.93 $\pm$ 2.01 & 84.91 $\pm$ 2.55 \\
    MusicGen & 300 M & 84.52 $\pm$ 2.19 & 85.11 $\pm$ 1.98\\
    MusicGen-MMD (ours) & 300 M & 84.18 $\pm$ 1.74 & \textbf{87.57 $\pm$ 2.16} \\

     \hline \bottomrule
    \end{tabular}
    }
    \label{tab:music-results-turk}
\end{table}

%% file: figures/figure_dac_mmd.tex
\pgfplotsset{
    every axis plot/.append style={line width=2pt},
    discard if not/.style 2 args={
        x filter/.append code={
            \edef\tempa{\thisrow{#1}}
            \edef\tempb{#2}
            \ifx\tempa\tempb
            \else
                \def\pgfmathresult{inf}
            \fi
        }
    }
}

\begin{figure*}
    \centering
\scalebox{0.97}{
\begin{tikzpicture}
  \begin{axis}[
    xlabel={MMD Weighting Factor},
    ylabel={MMD},
    xtick=data,
    xticklabels={$0$,$10^0$,$10^1$,$10^2$,$10^3$},
    name=mmd,
    width=\w,
    height=\h,
    xmode=log,
  ]
    \addplot+[blue, line width=0.1mm, mark options={fill=blue}] table[x=scale, y=mmd, col sep=comma]{mmdmidac.csv};
  \end{axis}

  \begin{axis}[
    xlabel={MMD Weighting Factor},
    ylabel={$\mathcal{I} (\%)$},
    xtick=data,
    xticklabels={$0$,$10^0$,$10^1$,$10^2$,$10^3$},
    name=mi,
    width=\w,
    height=\h,
    at={(mmd.south east)},
    xshift=0.075\textwidth,
    ylabel style={yshift=-0.15cm},
    xmode=log,
  ]
    \addplot+[orange, line width=0.1mm, mark options={fill=orange}] table[x=scale, y=mi, col sep=comma]{mmdmidac.csv};

  \end{axis}

  \begin{axis}[
    xlabel={MMD Weighting Factor},
    ylabel={MSSpec},
    xtick=data,
    xticklabels={$0$,$10^0$,$10^1$,$10^2$,$10^3$},
    name=msspec,
    width=\w,
    height=\h,
    at={(mi.south east)},
    ylabel style={yshift=-0.1cm},
    xshift=0.095\textwidth,
    xmode=log,
  ]
    \addplot+[black, line width=0.1mm, mark options={fill=black}] table[x=scale, y=msspec, col sep=comma, mark color=black]{mmdmidac.csv};

  \end{axis}
  
\end{tikzpicture}
}
    \caption{MMD, Mutual Information of RVQGAN \cite{kumar2024highfidelity} codes and MSSpec loss computed on our internal set. 
    MSSpec is the combination of $L1$ and $L2$ losses on the multi-resolution mel-spectrogram, used for reconstruction in EnCodec.
    The horizontal axis shows the weighting factor used for the MMD loss \textcolor{figc0}{$\mathcal{L}_\mathrm{inde}$}. 
     The total correlation $\mathcal{I}$ is computed on the whole 250k-samples training set for minimal bias in the histogram approximation. It is computed between two codebooks taken at random, averaged over five codebook couples, and expressed as a ratio to the entropy of the joint distribution (in \%).
    We removed the data point for the MMD weight of $10^4$ as the experiment diverged.}
    \label{fig:dac-mmd}
\end{figure*}

%% file: figures/figure_codecs_music.tex
\definecolor{cb1}{HTML}{D81B60}
\definecolor{cb2}{HTML}{1E88E5}
\definecolor{cb3}{HTML}{D29E02}
\definecolor{cb4}{HTML}{004D40}

\begin{figure}[h]
    \hspace{0cm}
\scalebox{.9}{
\begin{tikzpicture}
  \begin{axis}[
    xlabel={Codebook},
    ylabel={$\mathcal{I} (\%)$},
    yticklabels={$0$,$0$,$6 \cdot 10^{-2}$,$8 \cdot 10^{-2}$,$10 \cdot 10^{-2}$,$12 \cdot 10^{-2}$},
    xtick=data,
    name=mmd,
    width=0.4\textwidth,
    height=0.3\textwidth,
    legend columns=2,
    legend style={
    at={(xticklabel cs:.5)},
    anchor=north,
    xshift=-0.2cm,
    yshift=5.4cm,
},
  ]
    \addplot+[cb1, line width=0.1mm, mark options={fill=red}] table[x=codebook, y=mi, col sep=comma]{mi_encodec_24.csv};
    \addplot+[cb2, line width=0.1mm, mark options={fill=cb2}] table[x=codebook, y=mi, col sep=comma]{mi_encodec_32.csv};
    \addplot+[cb3, loosely dotted, line width=0.1mm, mark options={fill=cb3}] table[x=codebook, y=mi, col sep=comma]{mi_encodec_mmd_32.csv};
    \addplot+[cb4, dashed, line width=0.1mm, mark options={fill=cb4}] table[x=codebook, y=mi, col sep=comma]{mi_dac.csv};
    
  \legend{EnCodec-24kHz, EnCodec-32kHz, Encodec-32kHz-MMD, RVQGAN}
  \end{axis}

\end{tikzpicture}
}
    \vspace{-0.3cm}
    \caption{Mutual information between individual codebooks (on the horizontal axis) and all other codebooks, for difference codecs on FMA-Pop \cite{gui2024adapting}.}
    \label{fig:mi-fma-pop}
\end{figure}

%% file: figures/figure_codecs_speech.tex
\definecolor{cb1}{HTML}{D81B60}
\definecolor{cb2}{HTML}{1E88E5}
\definecolor{cb3}{HTML}{D29E02}
\definecolor{cb4}{HTML}{004D40}

\begin{figure}[h!]
\scalebox{0.9}{
\begin{tikzpicture}
    \hspace{0.1cm}
  \begin{axis}[
    xlabel={Codebook},
    ylabel={$\mathcal{I} (\%)$},
    xtick=data,
    yticklabels={$0$,$0$,$5 \cdot 10^{-2}$,$10 \cdot 10^{-2}$,$15 \cdot 10^{-2}$},
    name=mmd,
    width=0.4\textwidth,
    height=0.3\textwidth,
    legend columns=4,
    legend style={
    at={(xticklabel cs:.5)},
    anchor=north,
    xshift=-0.2cm,
    yshift=5.0cm,
},
  ]
    \addplot+[cb1, line width=0.1mm, mark options={fill=red}] table[x=codebook, y=mi, col sep=comma]{mi_encodec_24.csv};
    \addplot+[cb2, line width=0.1mm, mark options={fill=cb2}] table[x=codebook, y=mi, col sep=comma]{mi_speechtokenizer.csv};
    \addplot+[cb3, loosely dotted, line width=0.1mm, mark options={fill=cb3}] table[x=codebook, y=mi, col sep=comma]{mi_facodec.csv};
    
  \legend{EnCodec-24kHz, SpeechTokenizer, FACodec}
  \end{axis}

\end{tikzpicture}
}
    \vspace{-0.3cm}
    \caption{Mutual information between individual codebooks (on the horizontal axis) and all other codebooks, for difference codecs on LibriSpeech.}
    \label{fig:mi-speech}
\end{figure}

%% file: main.bbl
\begin{thebibliography}{51}
\providecommand{\natexlab}[1]{#1}
\providecommand{\url}[1]{\texttt{#1}}
\expandafter\ifx\csname urlstyle\endcsname\relax
  \providecommand{\doi}[1]{doi: #1}\else
  \providecommand{\doi}{doi: \begingroup \urlstyle{rm}\Url}\fi

\bibitem[Agostinelli et~al.(2023)Agostinelli, Denk, Borsos, Engel, Verzetti,
  Caillon, Huang, Jansen, Roberts, Tagliasacchi, Sharifi, Zeghidour, and
  Frank]{agostinelli2023musiclm}
Agostinelli, A., Denk, T.~I., Borsos, Z., Engel, J., Verzetti, M., Caillon, A.,
  Huang, Q., Jansen, A., Roberts, A., Tagliasacchi, M., Sharifi, M., Zeghidour,
  N., and Frank, C.
\newblock Music{LM}: Generating music from text.
\newblock \emph{arXiv preprint arXiv:2301.11325}, 2023.

\bibitem[Arjovsky et~al.(2017)Arjovsky, Chintala, and Bottou]{arjosky2014wgan}
Arjovsky, M., Chintala, S., and Bottou, L.
\newblock Wassertein generative adversarial networks.
\newblock \emph{Proc. Int. Conf. Machine Learning}, 2017.

\bibitem[Belghazi et~al.(2018)Belghazi, Baratin, Rajeswar, Ozair, Bengio,
  Courville, and Hjelm]{belghazi_mine_2021}
Belghazi, M.~I., Baratin, A., Rajeswar, S., Ozair, S., Bengio, Y., Courville,
  A., and Hjelm, R.~D.
\newblock {MINE}: {Mutual} {Information} {Neural} {Estimation}.
\newblock \emph{Proc. Int. Conf. Machine Learning}, 2018.

\bibitem[Borsos et~al.(2023)Borsos, Marinier, Vincent, Kharitonov, Pietquin,
  Sharifi, Roblek, Teboul, Grangier, Tagliasacchi, and
  Zeghidour]{borsos2023audiolm}
Borsos, Z., Marinier, R., Vincent, D., Kharitonov, E., Pietquin, O., Sharifi,
  M., Roblek, D., Teboul, O., Grangier, D., Tagliasacchi, M., and Zeghidour, N.
\newblock Audio{LM}: a language modeling approach to audio generation.
\newblock \emph{CoRR}, 2023.

\bibitem[Brakel \& Bengio(2017)Brakel and Bengio]{brakel2017learning}
Brakel, P. and Bengio, Y.
\newblock Learning independent features with adversarial nets for non-linear
  {ICA}.
\newblock \emph{Proc. Int. Conf. Machine Learning}, 2017.

\bibitem[Brown et~al.(2020)Brown, Mann, Ryder, Subbiah, Kaplan, Dhariwal,
  Neelakantan, Shyam, Sastry, Askell, Agarwal, Herbert-Voss, Krueger, Henighan,
  Child, Ramesh, Ziegler, Wu, Winter, Hesse, Chen, Sigler, Litwin, Gray, Chess,
  Clark, Berner, McCandlish, Radford, Sutskever, and Amodei]{brown2020language}
Brown, T.~B., Mann, B., Ryder, N., Subbiah, M., Kaplan, J., Dhariwal, P.,
  Neelakantan, A., Shyam, P., Sastry, G., Askell, A., Agarwal, S.,
  Herbert-Voss, A., Krueger, G., Henighan, T., Child, R., Ramesh, A., Ziegler,
  D.~M., Wu, J., Winter, C., Hesse, C., Chen, M., Sigler, E., Litwin, M., Gray,
  S., Chess, B., Clark, J., Berner, C., McCandlish, S., Radford, A., Sutskever,
  I., and Amodei, D.
\newblock Language models are few-shot learners.
\newblock \emph{Proc. Neural Inf. Process. Syst.}, 2020.

\bibitem[Burgess et~al.(2017)Burgess, Higgins, Pal, Matthey, Watters,
  Desjardins, and Lerchner]{burgess2018understanding}
Burgess, C.~P., Higgins, I., Pal, A., Matthey, L., Watters, N., Desjardins, G.,
  and Lerchner, A.
\newblock Understanding disentangling in $\beta$-{VAE}.
\newblock \emph{Proc. Neural Inf. Process. Syst.}, 2017.

\bibitem[Copet et~al.(2023)Copet, Kreuk, Gat, Remez, Kant, Synnaeve, Adi, and
  Défossez]{copet2023musicgen}
Copet, J., Kreuk, F., Gat, I., Remez, T., Kant, D., Synnaeve, G., Adi, Y., and
  Défossez, A.
\newblock Simple and controllable music generation.
\newblock \emph{Proc. Neural Inf. Process. Syst.}, 2023.

\bibitem[D{\'e}fossez et~al.(2023)D{\'e}fossez, Copet, Synnaeve, and
  Adi]{defossez2023high}
D{\'e}fossez, A., Copet, J., Synnaeve, G., and Adi, Y.
\newblock High fidelity neural audio compression.
\newblock \emph{Transactions on Machine Learning Research}, 2023.

\bibitem[Dhariwal et~al.(2020)Dhariwal, Jun, Payne, Kim, Radford, and
  Sutskever]{dhariwal2020jukebox}
Dhariwal, P., Jun, H., Payne, C., Kim, J.~W., Radford, A., and Sutskever, I.
\newblock Jukebox: A generative model for music.
\newblock \emph{arXiv preprint arXiv:2005.00341}, 2020.

\bibitem[Fan et~al.(2018)Fan, Lewis, and Dauphin]{fan-etal-2018-hierarchical}
Fan, A., Lewis, M., and Dauphin, Y.
\newblock Hierarchical neural story generation.
\newblock \emph{Proceedings of the 56th Annual Meeting of the Association for
  Computational Linguistics}, 2018.

\bibitem[Goodfellow et~al.(2014)Goodfellow, Pouget-Abadie, Mirza, Xu,
  Warde-Farley, Ozair, Courville, and Bengio]{goodfellow2014gan}
Goodfellow, I., Pouget-Abadie, J., Mirza, M., Xu, B., Warde-Farley, F., Ozair,
  S., Courville, A., and Bengio, Y.
\newblock Generative adversarial networks.
\newblock \emph{Proc. Neural Inf. Process. Syst.}, 2014.

\bibitem[Gray(1984)]{Gray1984VectorQ}
Gray, R.~M.
\newblock Vector quantization.
\newblock \emph{IEEE ASSP Magazine}, 1984.

\bibitem[Gretton et~al.(2012)Gretton, Bordwardt, Rasch, Schoelopf, and
  Smola]{mmd}
Gretton, A., Bordwardt, K., Rasch, M., Schoelopf, B., and Smola, A.
\newblock A kernel two-sample test.
\newblock \emph{Journal of Machine Learning Research}, 2012.

\bibitem[Gui et~al.(2024)Gui, Gamper, Braun, and
  Emmanouilidou]{gui2024adapting}
Gui, A., Gamper, H., Braun, S., and Emmanouilidou, D.
\newblock Adapting frechet audio distance for generative music evaluation.
\newblock In \emph{Proc. IEEE Int. Conf. Acoust. Speech Signal Process.}, 2024.
\newblock \doi{10.1109/ICASSP48485.2024.10446663}.

\bibitem[Higgins et~al.(2017)Higgins, Matthey, Pal, Burgess, Glorot, Botvinick,
  Mohamed, and Lerchner]{betavae}
Higgins, I., Matthey, L., Pal, A., Burgess, C., Glorot, X., Botvinick, M.,
  Mohamed, S., and Lerchner, A.
\newblock $\beta$-vae: Learning basic visual concepts with a constrained
  variational framework.
\newblock \emph{Proc. Int. Conf. Learning Repr.}, 2017.

\bibitem[Ho et~al.(2020)Ho, Jain, and Abbeel]{ho2020denoising}
Ho, J., Jain, A., and Abbeel, P.
\newblock Denoising diffusion probabilistic models.
\newblock \emph{Proc. Neural Inf. Process. Syst.}, 2020.

\bibitem[Huang et~al.(2023)Huang, Park, Wang, Denk, Ly, Chen, Zhang, Zhang, Yu,
  Frank, Engel, Le, Chan, Chen, and Han]{huang2023noise2music}
Huang, Q., Park, D.~S., Wang, T., Denk, T.~I., Ly, A., Chen, N., Zhang, Z.,
  Zhang, Z., Yu, J., Frank, C., Engel, J., Le, Q.~V., Chan, W., Chen, Z., and
  Han, W.
\newblock Noise2{M}usic: Text-conditioned music generation with diffusion
  models.
\newblock \emph{arXiv preprint arXiv:2302.03917}, 2023.

\bibitem[Huszar(2016)]{huszar_blogpost}
Huszar, F.
\newblock An alternative update rule for generative adversarial networks.
\newblock \emph{Blogpost}, 2016.

\bibitem[Hyvarinen et~al.(2023)Hyvarinen, Khemakhem, and
  Morioka]{hyvarinen_nonlinear_2023}
Hyvarinen, A., Khemakhem, I., and Morioka, H.
\newblock Nonlinear {Independent} {Component} {Analysis} for {Principled}
  {Disentanglement} in {Unsupervised} {Deep} {Learning}.
\newblock \emph{Patterns}, 2023.

\bibitem[ITU-R(2017)]{itu}
ITU-R.
\newblock Algorithms to measure audio programme loudness and true-peak audio
  level.
\newblock 2017.

\bibitem[Ju et~al.(2024)Ju, Wang, Shen, Tan, Xin, Yang, Liu, Leng, Song, Tang,
  Wu, Qin, Li, Ye, Zhang, Bian, He, Li, and Zhao]{ju2024naturalspeech}
Ju, Z., Wang, Y., Shen, K., Tan, X., Xin, D., Yang, D., Liu, Y., Leng, Y.,
  Song, K., Tang, S., Wu, Z., Qin, T., Li, X.-Y., Ye, W., Zhang, S., Bian, J.,
  He, L., Li, J., and Zhao, S.
\newblock Naturalspeech 3: Zero-shot speech synthesis with factorized codec and
  diffusion models.
\newblock In \emph{arXiv preprint arXiv:2403.03100}, 2024.

\bibitem[Juang \& Gray(1982)Juang and Gray]{juang1982multistage}
Juang, B.-H. and Gray, A.
\newblock Multiple stage vector quantization for speech coding.
\newblock \emph{Proc. IEEE Int. Conf. Acoust. Speech Signal Process.}, 1982.

\bibitem[Kharitonov et~al.(2022)Kharitonov, Lee, Polyak, Adi, Copet, Lakhotia,
  Nguyen, Rivière, Mohamed, Dupoux, and Hsu]{kharitonov2022textfree}
Kharitonov, E., Lee, A., Polyak, A., Adi, Y., Copet, J., Lakhotia, K., Nguyen,
  T.-A., Rivière, M., Mohamed, A., Dupoux, E., and Hsu, W.-N.
\newblock Text-free prosody-aware generative spoken language modeling.
\newblock \emph{Proceedings of the 60th Annual Meeting of the Association for
  Computational Linguistics}, 2022.

\bibitem[Kilgour et~al.(2019)Kilgour, Zuluaga, Roblek, and
  Sharifi]{kilgour2019frechet}
Kilgour, K., Zuluaga, M., Roblek, D., and Sharifi, M.
\newblock Fr\'echet audio distance: A metric for evaluating music enhancement
  algorithms.
\newblock \emph{INTERSPEECH}, 2019.

\bibitem[Kingma \& Welling(2014)Kingma and Welling]{kingma2014vae}
Kingma, D. and Welling, M.
\newblock Auto-encoding variational bayes.
\newblock \emph{Proc. Int. Conf. Learning Repr.}, 2014.

\bibitem[Kong et~al.(2020)Kong, Kim, and Bae]{kong2020hifigan}
Kong, J., Kim, J., and Bae, J.
\newblock Hifi-gan: Generative adversarial networks for efficient and high
  fidelity speech synthesis.
\newblock \emph{Proc. Neural Inf. Process. Syst.}, 2020.

\bibitem[Kong et~al.(2021)Kong, Ping, Huang, Zhao, and
  Catanzaro]{kong2021diffwave}
Kong, Z., Ping, W., Huang, J., Zhao, K., and Catanzaro, B.
\newblock Diffwave: A versatile diffusion model for audio synthesis.
\newblock \emph{Proc. Int. Conf. Learning Repr.}, 2021.

\bibitem[Koutini et~al.(2022)Koutini, Schlüter, Eghbal-zadeh, and
  Widmer]{Koutini_2022}
Koutini, K., Schlüter, J., Eghbal-zadeh, H., and Widmer, G.
\newblock Efficient training of audio transformers with patchout.
\newblock \emph{Proc. Interspeech}, 2022.

\bibitem[Kreuk et~al.(2023)Kreuk, Synnaeve, Polyak, Singer, Défossez, Copet,
  Parikh, Taigman, and Adi]{kreuk2023audiogen}
Kreuk, F., Synnaeve, G., Polyak, A., Singer, U., Défossez, A., Copet, J.,
  Parikh, D., Taigman, Y., and Adi, Y.
\newblock Audiogen: Textually guided audio generation.
\newblock \emph{Proc. Int. Conf. Learning Repr.}, 2023.

\bibitem[Kumar et~al.(2019)Kumar, Kumar, de~Boissiere, Gestin, Teoh, Sotelo,
  de~Brebisson, Bengio, and Courville]{kumar2019melgan}
Kumar, K., Kumar, R., de~Boissiere, T., Gestin, L., Teoh, W.~Z., Sotelo, J.,
  de~Brebisson, A., Bengio, Y., and Courville, A.
\newblock Melgan: Generative adversarial networks for conditional waveform
  synthesis.
\newblock \emph{Proc. Neural Inf. Process. Syst.}, 2019.

\bibitem[Kumar et~al.(2024)Kumar, Seetharaman, Luebs, Kumar, and
  Kumar]{kumar2024highfidelity}
Kumar, R., Seetharaman, P., Luebs, A., Kumar, I., and Kumar, K.
\newblock High-fidelity audio compression with improved rvqgan, 2024.

\bibitem[Li et~al.(2023)Li, YU, and Principe]{ddica}
Li, H., YU, S., and Principe, J.
\newblock Deep deterministic independent component analysis for hyperspectral
  unmixing.
\newblock \emph{Proc. IEEE Int. Conf. Acoust. Speech Signal Process.}, 2023.

\bibitem[Liu et~al.(2023{\natexlab{a}})Liu, Chen, Yuan, Mei, Liu, Mandic, Wang,
  and Plumbley]{liu2023audioldm}
Liu, H., Chen, Z., Yuan, Y., Mei, X., Liu, X., Mandic, D., Wang, W., and
  Plumbley, M.~D.
\newblock Audio{LDM}: Text-to-audio generation with latent diffusion models.
\newblock \emph{Proc. Int. Conf. Machine Learning}, 2023{\natexlab{a}}.

\bibitem[Liu et~al.(2023{\natexlab{b}})Liu, Tian, Yuan, Liu, Mei, Kong, Wang,
  Wang, Wang, and Plumbley]{liu2023audioldm2}
Liu, H., Tian, Q., Yuan, Y., Liu, X., Mei, X., Kong, Q., Wang, Y., Wang, W.,
  Wang, Y., and Plumbley, M.~D.
\newblock Audio{LDM} 2: Learning holistic audio generation with self-supervised
  pretraining.
\newblock \emph{arXiv preprint arXiv:2308.05734}, 2023{\natexlab{b}}.

\bibitem[Melechovsky et~al.(2023)Melechovsky, Guo, Ghosal, Majumder, Herremans,
  and Poria]{melechovsky2023mustango}
Melechovsky, J., Guo, Z., Ghosal, D., Majumder, N., Herremans, D., and Poria,
  S.
\newblock Mustango: Toward controllable text-to-music generation.
\newblock \emph{arXiv preprint arXiv:2311.08355}, 2023.

\bibitem[Radford et~al.(2019)Radford, Wu, Child, Luan, Amodei, and
  Sutskever]{radford2019language}
Radford, A., Wu, J., Child, R., Luan, D., Amodei, D., and Sutskever, I.
\newblock Language models are unsupervised multitask learners.
\newblock \emph{Technical Report}, 2019.

\bibitem[Raffel et~al.(2023)Raffel, Shazeer, Roberts, Lee, Narang, Matena,
  Zhou, Li, and Liu]{raffel2023exploring}
Raffel, C., Shazeer, N., Roberts, A., Lee, K., Narang, S., Matena, M., Zhou,
  Y., Li, W., and Liu, P.~J.
\newblock Exploring the limits of transfer learning with a unified text-to-text
  transformer.
\newblock \emph{Journal of Machine Learning Research}, 2023.

\bibitem[Ribeiro et~al.(2011)Ribeiro, Florêncio, Zhang, and
  Seltzer]{ribeiro2011crowdmos}
Ribeiro, F., Florêncio, D., Zhang, C., and Seltzer, M.
\newblock Crowdmos: An approach for crowdsourcing mean opinion score studies.
\newblock \emph{Proc. IEEE Int. Conf. Acoust. Speech Signal Process.}, 2011.

\bibitem[Rombach et~al.(2022)Rombach, Blattmann, Lorenz, Esser, and
  Ommer]{rombach2022highresolution}
Rombach, R., Blattmann, A., Lorenz, D., Esser, P., and Ommer, B.
\newblock High-resolution image synthesis with latent diffusion models.
\newblock \emph{Proc. IEEE/CVF Conf. Computer Vision and Pattern Recognition},
  2022.

\bibitem[Song \& Ermon(2019)Song and Ermon]{song2019generative}
Song, Y. and Ermon, S.
\newblock Generative modeling by estimating gradients of the data distribution.
\newblock \emph{Proc. Neural Inf. Process. Syst.}, 2019.

\bibitem[van~den Oord et~al.(2016)van~den Oord, Dieleman, Zen, Simonyan,
  Vinyals, Graves, Kalchbrenner, Senior, and Kavukcuoglu]{oord2016wavenet}
van~den Oord, A., Dieleman, S., Zen, H., Simonyan, K., Vinyals, O., Graves, A.,
  Kalchbrenner, N., Senior, A., and Kavukcuoglu, K.
\newblock Wavenet: A generative model for raw audio.
\newblock 2016.

\bibitem[Vasuki \& Vanathi(2006)Vasuki and Vanathi]{vasuki2006review}
Vasuki, A. and Vanathi, P.
\newblock A review of vector quantization techniques.
\newblock \emph{IEEE Potentials}, 2006.

\bibitem[Vaswani et~al.(2017)Vaswani, Shazeer, Parmar, Uszkoreit, Jones, Gomez,
  Kaiser, and Polosukhin]{vaswani2023attention}
Vaswani, A., Shazeer, N., Parmar, N., Uszkoreit, J., Jones, L., Gomez, A.~N.,
  Kaiser, L., and Polosukhin, I.
\newblock Attention is all you need.
\newblock \emph{Proc. Neural Inf. Process. Syst.}, 2017.

\bibitem[Villani(2009)]{villani2009ot}
Villani, C.
\newblock Optimal transport: Old and new.
\newblock \emph{Grundlehren der mathematischen Wissenschaften}, 2009.

\bibitem[Wang et~al.(2023)Wang, Chen, Wu, Zhang, Zhou, Liu, Chen, Liu, Wang,
  Li, He, Zhao, and Wei]{wang2023neural}
Wang, C., Chen, S., Wu, Y., Zhang, Z., Zhou, L., Liu, S., Chen, Z., Liu, Y.,
  Wang, H., Li, J., He, L., Zhao, S., and Wei, F.
\newblock Neural codec language models are zero-shot text to speech
  synthesizers.
\newblock \emph{arXiv preprint arXiv:2301.02111}, 2023.

\bibitem[Yang et~al.(2023{\natexlab{a}})Yang, Tian, Tan, Huang, Liu, Chang,
  Shi, Zhao, Bian, Wu, Zhao, Watanabe, and Meng]{yang2023uniaudio}
Yang, D., Tian, J., Tan, X., Huang, R., Liu, S., Chang, X., Shi, J., Zhao, S.,
  Bian, J., Wu, X., Zhao, Z., Watanabe, S., and Meng, H.
\newblock Uniaudio: An audio foundation model toward universal audio
  generation.
\newblock \emph{arXiv preprint arXiv:2310.00704}, 2023{\natexlab{a}}.

\bibitem[Yang et~al.(2023{\natexlab{b}})Yang, Yu, Wang, Wang, Weng, Zou, and
  Yu]{yang2023diffsound}
Yang, D., Yu, J., Wang, H., Wang, W., Weng, C., Zou, Y., and Yu, D.
\newblock Diffsound: Discrete diffusion model for text-to-sound generation.
\newblock \emph{IEEE/ACM Trans. Audio Speech Lang. Process.},
  2023{\natexlab{b}}.

\bibitem[Yu et~al.(2021)Yu, Alesiani, Yu, Jenssen, and
  Principe]{yu_measuring_2021}
Yu, S., Alesiani, F., Yu, X., Jenssen, R., and Principe, J.~C.
\newblock Measuring {Dependence} with {Matrix}-based {Entropy} {Functional}.
\newblock \emph{AAAI}, 2021.

\bibitem[Zeghidour et~al.(2021)Zeghidour, Luebs, Omran, Skoglund, and
  Tagliasacchi]{zeghidour2021soundstream}
Zeghidour, N., Luebs, A., Omran, A., Skoglund, J., and Tagliasacchi, M.
\newblock Sound{S}tream: An end-to-end neural audio codec.
\newblock \emph{arXiv preprint arXiv:2107.03312}, 2021.

\bibitem[Zhang et~al.(2024)Zhang, Zhang, Li, Zhou, and
  Qiu]{zhang2024speechtokenizer}
Zhang, X., Zhang, D., Li, S., Zhou, Y., and Qiu, X.
\newblock Speechtokenizer: Unified speech tokenizer for speech large language
  models.
\newblock In \emph{Proc. Int. Conf. Learning Repr.}, 2024.

\end{thebibliography}
